\documentclass[sigconf]{acmart}

\AtBeginDocument{%
  \providecommand\BibTeX{{%
    \normalfont B\kern-0.5em{\scshape i\kern-0.25em b}\kern-0.8em\TeX}}}

\setcopyright{acmlicensed}

\copyrightyear{2025}
\acmYear{2025}
\setcopyright{cc}
\setcctype{by}
\acmConference[CHI '25]{CHI Conference on Human Factors in Computing Systems}{April 26-May 1, 2025}{Yokohama, Japan}
\acmBooktitle{CHI Conference on Human Factors in Computing Systems (CHI '25), April 26-May 1, 2025, Yokohama, Japan}\acmDOI{10.1145/3706598.3713337}
\acmISBN{979-8-4007-1394-1/25/04}

\usepackage{multirow}
\usepackage{enumitem}

\usepackage{listings}
\usepackage{xcolor}

\begin{document}

\title[Generative AI in Knowledge Work: Design Implications for Data
Navigation and Decision-Making]{Generative AI in Knowledge Work: Design Implications for Data
Navigation and Decision-Making}


\author{Bhada Yun}
\authornote{Yun and Feng contributed equally to this work (order was decided by coin toss).}
\affiliation{%
  \institution{University of California, Berkeley}
  \city{Berkeley}
  \state{CA}
  \country{USA}
}
\email{bhadayun@berkeley.edu}

\author{Dana Feng}
\authornotemark[1]
\affiliation{%
  \institution{University of California, Berkeley}
  \city{Berkeley}
  \state{CA}
  \country{USA}
}
\email{danafeng@berkeley.edu}

\author{Ace S. Chen}
\affiliation{%
  \institution{University of California, Berkeley}
  \city{Berkeley}
  \state{CA}
  \country{USA}
}
\email{12aschen@berkeley.edu}

\author{Afshin Nikzad}
\affiliation{%
  \institution{University of Southern California}
  \city{Los Angeles}
  \state{CA}
  \country{USA}
}
\email{afshinni@usc.edu}

\author{Niloufar Salehi}
\affiliation{%
 \institution{University of California, Berkeley}
 \city{Berkeley}
 \state{CA}
 \country{USA}
}
\email{nsalehi@berkeley.edu}

\renewcommand{\shortauthors}{Yun and Feng et al.}



\begin{abstract}
Our study of 20 knowledge workers revealed a common challenge: the difficulty of synthesizing unstructured information scattered across multiple platforms to make informed decisions. Drawing on their vision of an ideal knowledge synthesis tool, we developed Yodeai, an AI-enabled system, to explore both the opportunities and limitations of AI in knowledge work. Through a user study with 16 product managers, we identified three key requirements for Generative AI in knowledge work: adaptable user control, transparent collaboration mechanisms, and the ability to integrate background knowledge with external information. However, we also found significant limitations, including overreliance on AI, user isolation, and contextual factors outside the AI's reach. As AI tools become increasingly prevalent in professional settings, we propose design principles that emphasize adaptability to diverse workflows, accountability in personal and collaborative contexts, and context-aware interoperability to guide the development of human-centered AI systems for product managers and knowledge workers.
\end{abstract}

\ccsdesc[500]{Human-centered computing~Empirical studies in HCI}
\ccsdesc[300]{Human-centered computing~Empirical studies in interaction design}

\keywords{Knowledge Synthesis, Information Visualization, Human-AI Interaction, Large Language Models, Interaction Design}



\maketitle
\section{Introduction}

\begin{quote}
 ``Managers are not confronted with problems that are independent of each other, but with dynamic situations that consist of complex systems of changing problems that interact with each other. I call such situations, messes.'' \\- Russell Ackoff, 1979
\end{quote}

According to some estimates, 80\% of the data in organizations is messy, unstructured, and rarely used \cite{sint2009combining}. Knowledge workers face challenges in handling the scale and complexity of unstructured data, which leads to difficulties in extracting reliable information and making informed decisions \cite{marshall2008rethinking, marshall2008rethinking2, whittaker1996email, bergman2003user, jones2007personal, whittaker2001character}. AI-based tools have the potential to address these challenges by automatically structuring and synthesizing information \cite{zheng2023large, pan2024unifying}. However, AI models are not well equipped to take over the entire data analysis process, as it is often an iterative and collaborative process that relies on the domain knowledge of the decision maker \cite{suchman1987plans}. Furthermore, the potential for hallucinations in the model \cite{leiser2023chatgpt} and biases \cite{weidinger2022taxonomy} raise concerns about the validity of purely AI-generated outputs. As companies increasingly integrate AI into their employees' workflows, it is important to understand what opportunities and limitations generative AI poses for data navigation and decision-making in knowledge work.

In this paper, we build an AI-enabled system for product managers (PMs) as a proxy for exploring the experience and
opinions of knowledge workers with generative AI. Prior literature has defined the use of proxy systems as utilizing
existing technologies to ideate for future technologies \cite{pierson2006walking}; we follow a similar pattern, developing an AI system
to glean principles for future knowledge work tools. Our formative study with 20 knowledge workers, including PMs, consultants, journalists, and graduate students, revealed three critical challenges in their work: searching and synthesizing vast amounts of unstructured data, extracting reliable insights, and maintaining effective collaboration. Knowledge workers struggle to synthesize information from a variety of sources, ranging from user interviews to research articles \cite{hniche2023entreprise, denning2006infoglut}. This challenge is compounded by the tacit nature of domain expertise required to effectively analyze and compare data \cite{chen2018personal, kulesza2013mentalmodels}. We found that successful AI tools for human AI collaborative knowledge synthesis must integrate work context, personal domain knowledge, and the iterative nature of decision-making \cite{pedrycz2015iterativeness}.

We designed Yodeai to integrate into existing knowledge workflows by enabling knowledge workers to explore and synthesize unstructured data (e.g., product reviews and customer interview transcripts) through interactive widgets, customizable workflow templates that transform unstructured data into structured insights. To empirically examine knowledge workers' experiences with AI-enabled tools, we conducted an 80-minute study with a group of 16 product managers. Participants role-played as PMs at Notion, a software company, analyzing two data collections (12 YouTube interview transcripts and 132 App Store reviews) using three interactive widgets (User Insights, Pain Point Tracker, and Q\&A). The study involved a divergent phase where participants brainstormed six new feature or bug fix proposals, followed by a convergent phase where they prioritized their top three proposals. Over 20 minutes of system onboarding, 25 minutes of task work, and pre/post task interviews and surveys, we gathered rich insights about knowledge workers' interactions with and perspectives on AI-enabled knowledge synthesis.

Our results show a need for AI knowledge work tools to support diverse data exploration and decision-making workflows, enabling participants to leverage tool components in unique ways that suit their individual needs and preferences. With Yodeai, we show that the flexibility of the system allowed users to integrate their own domain expertise with the AI's outputs, as well as verify their own and others' work. However, our findings also highlighted potential challenges, such as overreliance on automation, navigating stale data, and privacy concerns.

Based on these findings, we propose three key design implications for developing effective and adaptable AI interactions for complex knowledge work: adaptability, accountability, and interoperability. Our work reveals a deeper understanding of the use of AI in knowledge work, as well as the challenges and opportunities that AI presents. To summarize, our contributions are the following:
\begin{itemize}
\item{Specific challenges knowledge workers face in navigating data through organization and structure, collaboration in terms of context and integrations, and decision making with prioritization and search.}
\item Empirical evaluation of the opportunities and limitations of generative AI for complex knowledge work with Yodeai, an AI tool that supports data exploration, verification, prioritization, and collaboration.
\item Design implications for AI in knowledge work that depends on understanding and reasoning over data, such as product management (Table \ref{tab:AI_Design_Principles}), emphasizing adaptability to diverse workflows, personal and collaborative accountability, and context-aware interoperability.
\end{itemize}
\section{Background And Related Work}

\subsection{AI use in Knowledge Work}

The integration of AI, such as generative models, has led to productivity gains in multiple domains of knowledge work \cite{bubeck2023sparks, eloundou2023gpts}, including management consulting \cite{dell2023navigating}, software development \cite{welsh2022end, qian2024take}, and writing \cite{noy2023experimental}. AI can assist writers with tasks such as planning, translation, and proofreading \cite{gero2023social}, while UX designers benefit from AI-generated design briefs and filler text for mockups \cite{li2023user}. In the medical field, AI has been applied to rehabilitation assessment \cite{lee2021human} and sepsis treatment \cite{sendak2020human}. A 25-year survey revealed that AI can assist with general human resource management tasks, such as recruitment, interviews, and employee training \cite{pereira2023systematic}, while in academia, interviews demonstrated the use of LLM in HCI research \cite{kapania2024m}. 

In addition to investigating new ways of using generative AI for specific knowledge work tasks, existing literature has also focused on interviews with knowledge workers about their experiences with and perceptions of AI. Research workshops conducted by Woodruff et al. \cite{woodruff2023knowledge} in industries of knowledge work such as advertising, law, and software development revealed that many workers believed that generative AI can automate menial work, but that a human in the loop is needed to ensure a certain standard is met. The workshops also revealed that AI could cause issues of dehumanization, as AI lacks the ability to perform interpersonal work, disconnection, where the work produced lacks social awareness and originality, and disinformation, where hallucinations occur. A large-scale survey conducted by Brachman et al. \cite{brachman2024knowledge} revealed how knowledge workers utilize LLM in their work, as well as how they would like to use LLMs for creation, information, advice, and automation. Finally, a two-week diary study by Kobiella et al. on young professionals under the age of 30 further emphasizes the issue of disconnect, where workers feel a lack of challenge, ownership and quality in LLM outputs, as well as deskilling, where barriers to information synthesis have been reduced \cite{kobiella2024if}. 

We build on this work and develop an interactive system, Yodeai, to observe how knowledge workers interact with AI tools to navigate information and make decisions in practice. With this system, we are able to identify practical insights supported by direct hands-on experience, extending beyond studies that only use surveys or visual demonstrations \cite{brachman2024knowledge, woodruff2023knowledge, li2024user}. Through this approach, we contribute design principles for future AI systems in knowledge work.


\subsection{AI-Assisted Data Exploration and Collaboration}

HCI research has long been invested in studying and developing new models and interactions for information seeking and synthesis \citep{wilson1999models, whittaker2011personal, jones2007personal, jones2010keeping, hearst2009search, epstein2020mapping, li2010stage, boardman2004stuff, whittaker2006email, bernstein2008information}. Knowledge workers often rely on external representations such as file systems or information `boxes' to navigate the vast scale of information at different levels of abstraction \cite{crescenzi2021supporting}. To help understand external and personal data, various interfaces have been developed using LLMs. These interfaces employ a wide range of output formats, from whiteboards \cite{palani2022interweave, suh2023sensecape}, hierarchies \cite{suh2023sensecape, kang2023synergi}, summarization \cite{dang2022beyond, liu2023selenite}, graphical structures \cite{jiang2023graphologue, li2023meddm}, to box structures \cite{crescenzi2021supporting}. 

Researchers have developed tools to support the navigation of exploratory data, both external (e.g., publicly available information from the Internet) and personal data (e.g., transcribed interviews). Synergi generates LLM-based summaries for papers cited in academic articles \cite{kang2023synergi}, while Selenite combines web and LLM-generated content to offer an overview of options and criteria for activities such as shopping \cite{liu2023selenite}. In the realm of personal/collaborative data exploration, numerous LLM tools have been created across disciplines such as data science and writing \cite{hassan2023chatgpt, dang2022beyond}. 
Recent developments like GPT plugins \cite{gptplugins} and personalized GPTs \cite{gpts} have highlighted the challenge of integrating personal data and specialized actions into AI systems. Building on insights from personal LLM agents that handle data analysis, developer APIs, and web integrations \cite{xie2023openagents}, our system combines web-sourced information (e.g. Apple App Store reviews) with user's personal files to support comprehensive knowledge synthesis.

Prior research has demonstrated various approaches to help users explore LLM outputs more effectively. For example, Sensecape \cite{suh2023sensecape} facilitates non-linear tasks like trip planning through text extraction and semantic exploration features, while ExploreLLM \cite{ma2023beyond} enables structured thinking and exploration of options for tasks such as restaurant discovery. Building on these advances, Yodeai introduces interactive widgets that allow users to engage with LLM outputs through multiple modalities—conversational, visual, and quantitative—to deepen their understanding and exploration of AI-generated insights.

The design of human-AI interactions presents unique challenges due to AI's uncertain capabilities and variable outputs \cite{yang2020re}. Researchers have addressed these challenges through specialized LLM frameworks for data exploration, such as decomposing LLM generation into input units, model instances, and output spaces \cite{kim2023cells}, and using graphical models to break down tasks for improved LLM prompting \cite{besta2023graph}. Recent work has proposed comprehensive frameworks for human-AI collaboration, including multimodal interaction approaches \cite{shi2023understanding}, design principles for generative AI applications \cite{weisz2024design}, and guidelines for AI interactions across different stages, from before use, during use, when something goes wrong, and over time \cite{amershi2019guidelines}. While these frameworks provide valuable theoretical foundations, our work extends beyond theory by implementing a prototype AI system, allowing us to derive concrete design implications from an empirical study with a focused group of knowledge workers that inform the design of more human-centric future knowledge work tools.

\subsection{AI Supported Decision-Making and Collaboration}

HCI research has produced diverse decision support tools for both personal and professional contexts \cite{yun2021human, rundo2020recent, rathnam1995tools}. Personal tools such as Mesh \cite{chang2020mesh} and ProactiveAgent \cite{ma2023proactiveagent} facilitate faster problem solving through techniques such as comparison tables and task decomposition \cite{yen2023coladder, liu2019unakite}. In the professional sphere, specialized tools have emerged to address domain-specific needs: ParaLib for library comparison \cite{yan2022concept}, Threddy for research analysis \cite{kang2022threddy}, Luminate for creative decision-making \cite{suh2023structured}, and SepsisLab for medical diagnosis \cite{zhang2023rethinking}. Building on these academic advances and industry solutions like Dovetail's automatic thematic clustering and Productboard's feature request prioritization, we tailored our system specifically for product management workflows.

The effectiveness of AI-assisted decision-making depends on a complex interplay of factors related to the human decision maker, the task at hand, and AI itself \cite{eigner2024determinants}. Literature surveys \cite{lai2021towards} and user behavioral studies \cite{munyaka2023decision, yen2024search} have examined human-AI decision-making tasks, evaluation metrics, and the impact of AI partners on human decision-making. This literature emphasizes the nuances of joint decision-making, as user background knowledge can affect the problem-solving process and results \cite{inkpen2023advancing}, with trust and explainability playing key roles in the adoption of generative AI applications \cite{amershi2019guidelines}. Our study extends this understanding by examining these dynamics specifically within product management—a domain where decisions often require synthesis of diverse user needs, technical constraints, and business priorities. In doing so, we uncover patterns that apply to knowledge work context where workers must synthesize unstructured information, balance competing priorities, and make evidence-based decisions.
\section{Formative Study}

We conducted semi-structured remote interviews with 20 participants (E1-E20) located in the United States. To ensure diverse perspectives on knowledge work and information management, we recruited participants across various professional roles: 2 journalists (E1, E18), 5 startup founders (E2, E6, E7, E9, E16), 3 graduate students (E3-E5), 2 consultants (E8, E11), 7 product managers (E10, E12-E15, E19, E20), and 1 associate director (E17). Each interview lasted 45-60 minutes and participants were compensated {\verb|$20|}. 

\begin{table*}
\centering
\caption{Participant Overview and Background}
\includegraphics[width=0.95\linewidth]{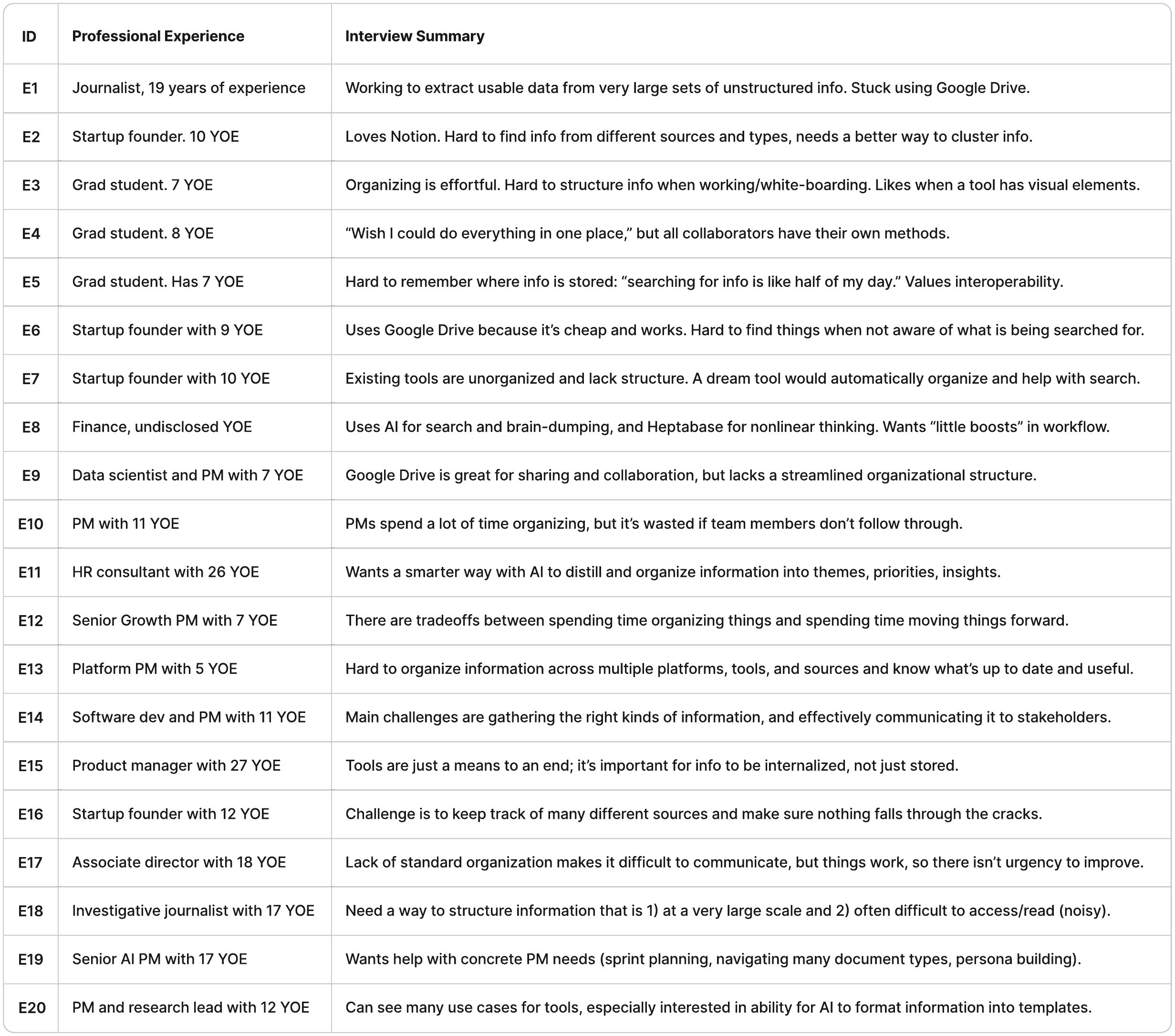}
\Description{A table showing interview summaries from 20 professionals for the formative study, labeled E1 through E20. E1 is a Journalist with 19 years experience working to extract usable data from very large sets of unstructured info and stuck using Google Drive. E2 is a Startup founder with 10 YOE who loves Notion but finds it hard to find info from different sources and types, needing a better way to cluster info. E3 is a Grad student with 7 YOE who finds organizing effortful and structuring info when working/white-boarding hard, but likes when a tool has visual elements. E4 is a Grad student with 8 YOE who wishes they could do everything in one place but notes all collaborators have their own methods. E5 is a Grad student with 7 YOE who finds it hard to remember where info is stored, spending half their day searching, and values privacy and interoperability. E6 is a Startup founder with 9 YOE using Google Drive for its affordability and functionality but struggles to find things when unsure what to search for. E7 is a Startup founder with 10 YOE who finds existing tools unorganized and lacking structure, wanting automatic organization and search help. E8 is in Finance with undisclosed experience, uses AI for search and brain-dumping plus Heptabase for nonlinear thinking, wanting workflow boosts. E9 is a Data scientist and PM with 7 YOE who likes Google Drive for sharing and collaboration but finds it lacks streamlined organization. E10 is a PM with 11 YOE noting PMs spend much time organizing but it's wasted if team members don't follow through. E11 is an HR consultant with 26 YOE wanting smarter AI-driven information organization into themes, priorities, and insights. E12 is a Senior Growth PM with 7 YOE discussing tradeoffs between organizing and moving things forward. E13 is a Platform PM with 5 YOE struggling to organize information across multiple platforms and determine what's current. E14 is a Software dev and PM with 11 YOE facing challenges in gathering and communicating information to stakeholders. E15 is a Product manager with 27 YOE emphasizing tools are means to ends and information needs internalization. E16 is a Startup founder with 12 YOE challenged by tracking various sources without missing anything. E17 is an Associate director with 18 YOE noting lack of standard organization hampers communication but status quo works adequately. E18 is an Investigative journalist with 17 YOE needing to structure large-scale, often noisy information. E19 is a Senior AI PM with 17 YOE seeking help with concrete PM needs like sprint planning and persona building. E20 is a PM and research lead with 12 YOE interested in AI's potential for formatting information into templates.}
\label{tab:AI_Design_Principles}
\end{table*}

\subsection{Interview Protocol Development}

Our semi-structured interviews used an exploratory approach to uncover knowledge workers' information management practices, challenges, and needs. The flexible format allowed participants to elaborate on topics most significant to them. For each section outlined below, we prepared a list of potential open-ended questions to select from as the conversation unfolded. The interviews were structured into three main stages:

\subsubsection{Information Management}

We started each interview by exploring how participants personally manage their information, seeking specific anecdotes and insights into the types of information they worked with as well as the tools they have used. Example questions include:

\begin{itemize}
  \item Is there a specific type of information (e.g. medical records) that you find particularly challenging to organize?
  \item Can you tell me about how you organize your information? What tools/apps do you use?
  \item Have you tried other solutions or methods to organize your information in the past? What worked and what didn't?
\end{itemize}

\subsubsection{Collaboration Dynamics}

Next, we explored how their information management practices influenced their work with others, given the collaborative nature of knowledge work. Example questions include:

\begin{itemize}
  \item How do you share information with others (e.g. colleagues)? Are there challenges with your current method?
  \item Do you collaborate with others on documents or files? If so, what challenges do you face in this process?
  \item Can you tell me about times you needed to capture new information quickly? Where do you encounter times like this the most—work/personal/team?
\end{itemize}

\subsubsection{Future Tool Perspectives}

The interviews concluded by exploring participants' tool needs as well as their attitudes toward AI assistance. Example questions include:

\begin{itemize}
  \item Are there certain features or capabilities you wish existed in your current tools that would make organizing information easier for you?
  \item What would an ideal solution look like for you?
  \item How do you feel about using AI or automation in helping you organize or find information?
\end{itemize}

\subsection{Data Analysis}

To analyze the data, we employed line-by-line coding of each interview transcript, with a primarily top-down approach \cite{bingham2021deductive} using predefined categories of ``Organization'', ``Search'', ``Collaboration'', and ``Tool Needs and Wants'' that was developed after conducting the interviews. Next, we identified common subthemes across the codes, including ``information relationships'' or ``getting context'' using affinity mapping \cite{harboe2015real}. In the next section, we describe our findings from this formative study that led to the core design thesis of Yodeai and our follow up user study: enabling efficient and collaborative data navigation and decision-making.

\subsection{Data Navigation: Organization \& Structure}

\paragraph{General remarks:}
Our analysis revealed that the challenges of information organization varied significantly depending on the use of the tool and the context of work. Google Drive users, particularly in startup environments that ``move fast'' (E7, E9), reported specific difficulties with their information organization systems. These challenges included frequent disorganization due to inconsistent folder structures, unclear file naming conventions, and the tendency to create individual bloated documents for all meetings, leading to priorities being missed or overlooked (E2, E4, E9). Meanwhile, Notion users, while appreciating the tool's flexibility, noted challenges with feature overwhelm and too many templates (E3). Participants also emphasized that they work with both quantitative (sales metrics) and qualitative data (transcripts) (E13-14), which require different methods of analysis.

\paragraph{Tool wants and needs}

Participants voiced the need for a centralized platform that could automatically infer an organizational structure and provide intuitive search functionality (E7, E10, E11, E13). For instance, E2 wanted a tool that could help ``organize things while I'm on the go'' while E10 shared that they would want a tool that gives snapshots of information and intelligently indexes the information. P20 also noted the importance of standardized templates, and the wish to have AI produce things in template forms as they already use Google suite templates.

\subsection{Collaboration: Context \& Integrations}

\paragraph{General remarks:}

Corporate knowledge workers (17 out of 20 participants) emphasized frequent information sharing within their companies. While E12 stressed the importance of organized information for team accessibility, E17 highlighted that there is no standard operating procedure on how to share and manage information, and that people do whatever works for them in terms of tool choice. E6 further emphasized the time sink of manual sharing, stating ``I’m stuck, I literally need to spend an hour—no—a day and a half sifting through all these docs, tagging people, adding them.'' Additionally, E12 noted that a lack of context makes it hard to identify the right stakeholders, while E14 highlighted challenges in communicating results to stakeholders who ``don't speak the same language.''

\paragraph{Tool wants and needs:}

Participants wanted tools that would allow them to understand the knowledge bases of their colleagues without requiring real-time communication (``offline context-gaining'') (E2). E13 noted that a tool that could pull in notes and meeting transcripts would be very important. Integration with existing platforms (e.g. Salesforce for customer data) was deemed crucial for cross-team collaboration (E3, E5, E7-E9, E13, E20). E6 emphasized the need for a tool that could handle collaboration ``when it's other people's content that is interlaced with my workflow''.

\subsection{Decision Making: Prioritization \& Search}

\paragraph{General remarks:}

Participants expressed the need for tools that support the analysis, ideation, and prioritization of data, revealing the various divergent and convergent aspects of decision making (E11-E13, E17, E18). They reported difficulties in finding previously known information (E2, E9, E16). E5 described information search as ``half of my day'' while E16 explained that ``finding info is hard, opening ten documents to find the one thing you were looking for.'' Many found keyword search limited, only useful when sufficient information context is known, which can be troublesome when sharing data repositories with others who lack familiarity with the data (E2, E6, E7, E12).

\paragraph{Tool wants and needs:}

Participants wanted a visualization that helps them understand content relationships (E12) and recognized the potential in an AI assistant that could help ``prioritize ideas, and give hotspots or find themes for me'' (E11). They emphasized the need for semantic search, thematic clustering, and information visualization (E2, E9, E12). E12 and E14 expressed the desire for an AI assistant to take notes and generate action items and meeting summaries, while E13 preferred a system capable of independently identifying insights and patterns without relying on user-prompted questions. Importantly, many participants (E6, E9, E10, E15)  emphasized showcasing sources, as ``people want to be able to know what the authenticity of the document is'' (E1).

\subsection{Additional Findings}

While this paper focuses on the formative study's interview findings that are most relevant to our system design, our study generated additional insights about domain-specific challenges outside of product management, such as in journalism or startup customer discovery. These findings have informed our broader research agenda and will be explored in future work focusing on enterprise knowledge management practices and AI-assisted collaboration.
\section{Yodeai}

\begin{figure*} [h!]
  \centering
  \includegraphics[width=\textwidth]{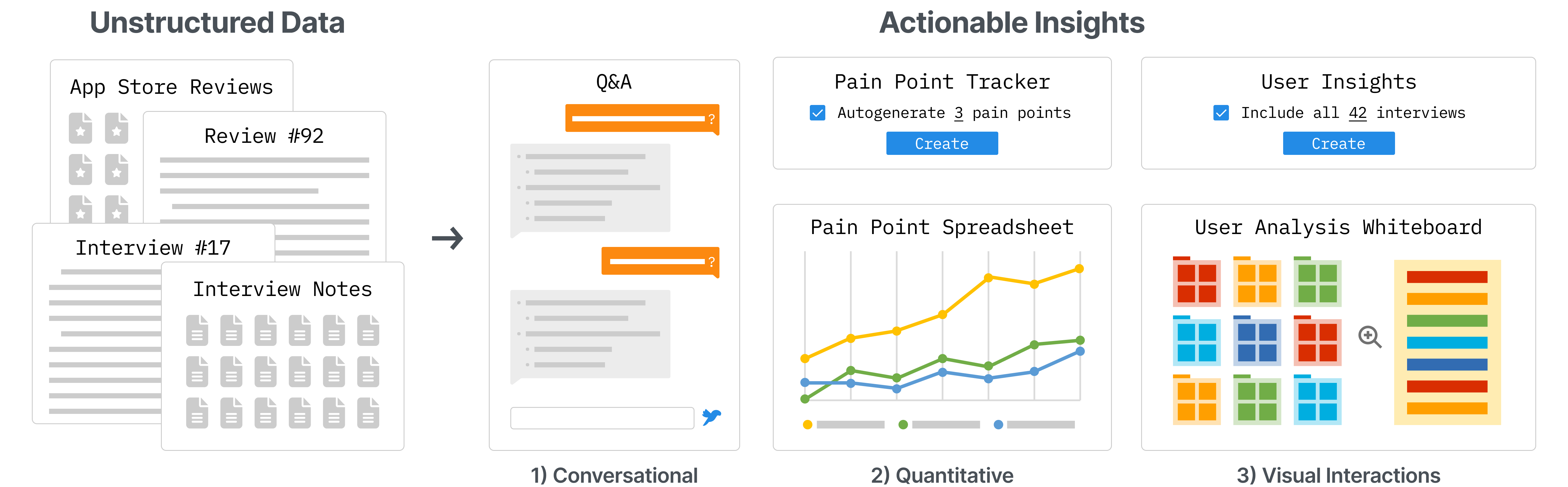}
  \caption{Yodeai hosts three AI-powered workflow templates (widgets) that transform unstructured data into actionable insights. 1) Q\&A: ask questions and retrieve relevant information with citations (conversational). 2) Pain Point Tracker: cluster and quantify key issues over time (quantitative). 3) User Insights: view multi-granular summaries on a whiteboard (visual). With these widgets, users can explore data, identify key points, and visualize metrics to support exploration and decision-making.}
  \Description{Yodeai transforms unstructured data into actionable insights through interactive widgets. The image on the left has a label of ``Unstructured Data'', with app store reviews and interviews, and an arrow points to the image on the right labeled ``Actionable Insights'' where there are 3 key components, a Q\&A widget that is conversational, a Pain Point Tracker that allows for auto-generation of pain points and plots a spreadsheet over time, allowing for quantitative use, and a User Insights widget where users can specify which reviews are included and automatically create a whiteboard of sticky notes per user and a high level summary, allowing for visual interactions.}
\end{figure*}

To investigate generative AI's potential in collaborative data navigation and decision-making, we developed Yodeai, an AI system focusing on PMs as a representative case of knowledge workers. We chose PMs because they exemplify the complexity of modern knowledge work—they regularly synthesize diverse data sources, collaborate across teams, and make data-driven decisions. Furthermore, our formative study included a large number of PM or PM adjacent experiences.

\subsection{Design Process}
\begin{figure}[h!]
    \centering
    \begin{minipage}{0.45\textwidth}
        \centering
        \includegraphics[width=\linewidth]{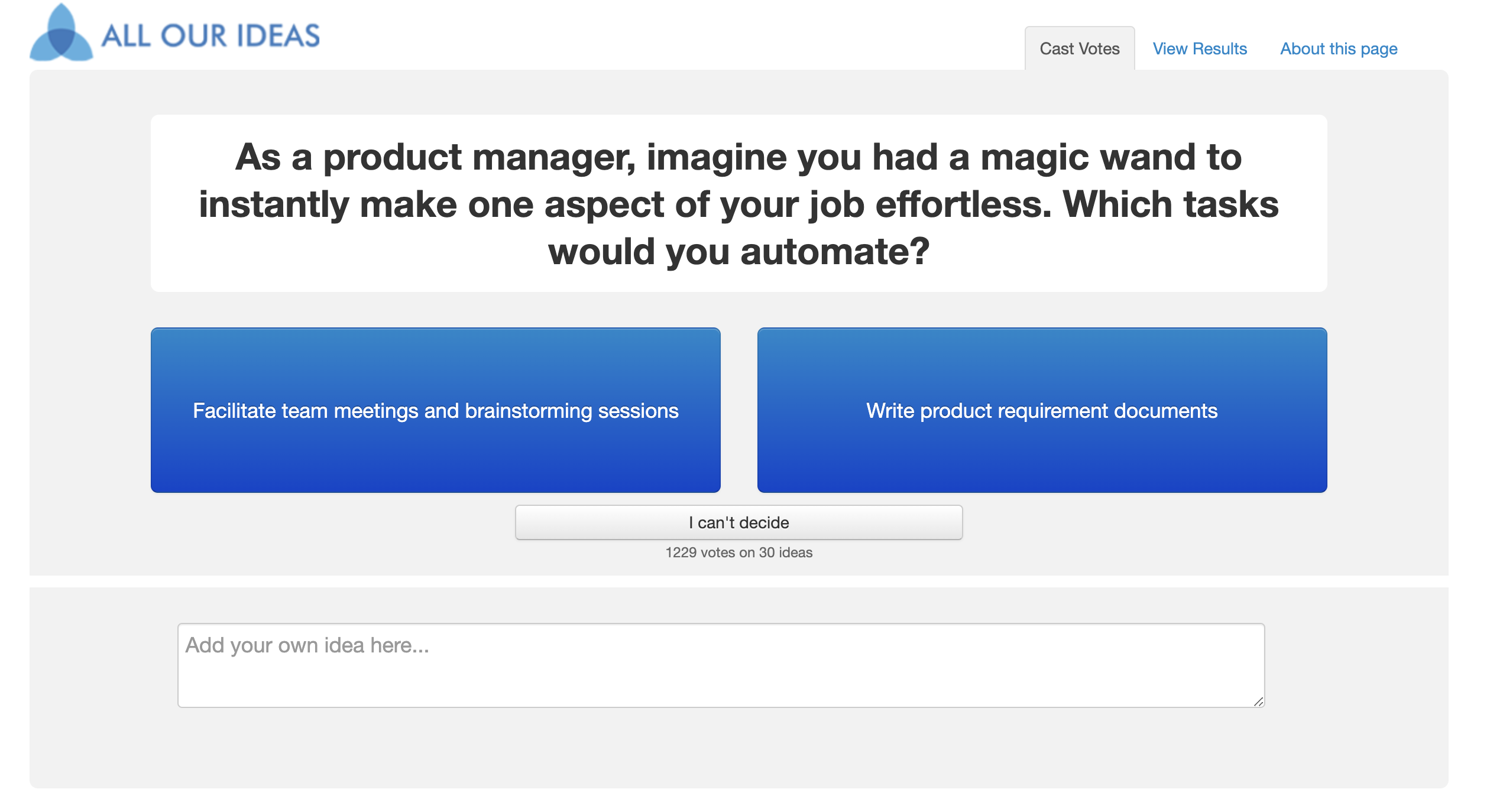}
        \caption{A voting interface for PMs to either submit their own ideas in response to the prompt or vote on existing ideas for PM tasks that AI could potentially assist with.}
        \Description{The figure showcases the All Our Ideas website where the poll was conducted on, and has the question ``As a PM, imagine you had a magic wand to instantly make one aspect of your job effortless. Which tasks would you automate?''. Underneath the question showcases two choices, ``Extract and summarize user pain points from feedback and reviews'', as well as ``Monitor and respond to user feedback on social media and forums''.}
        \label{fig:magicwand1}
    \end{minipage}
    \hfill  
    \begin{minipage}{0.45\textwidth}
        \centering
        \includegraphics[width=\linewidth]{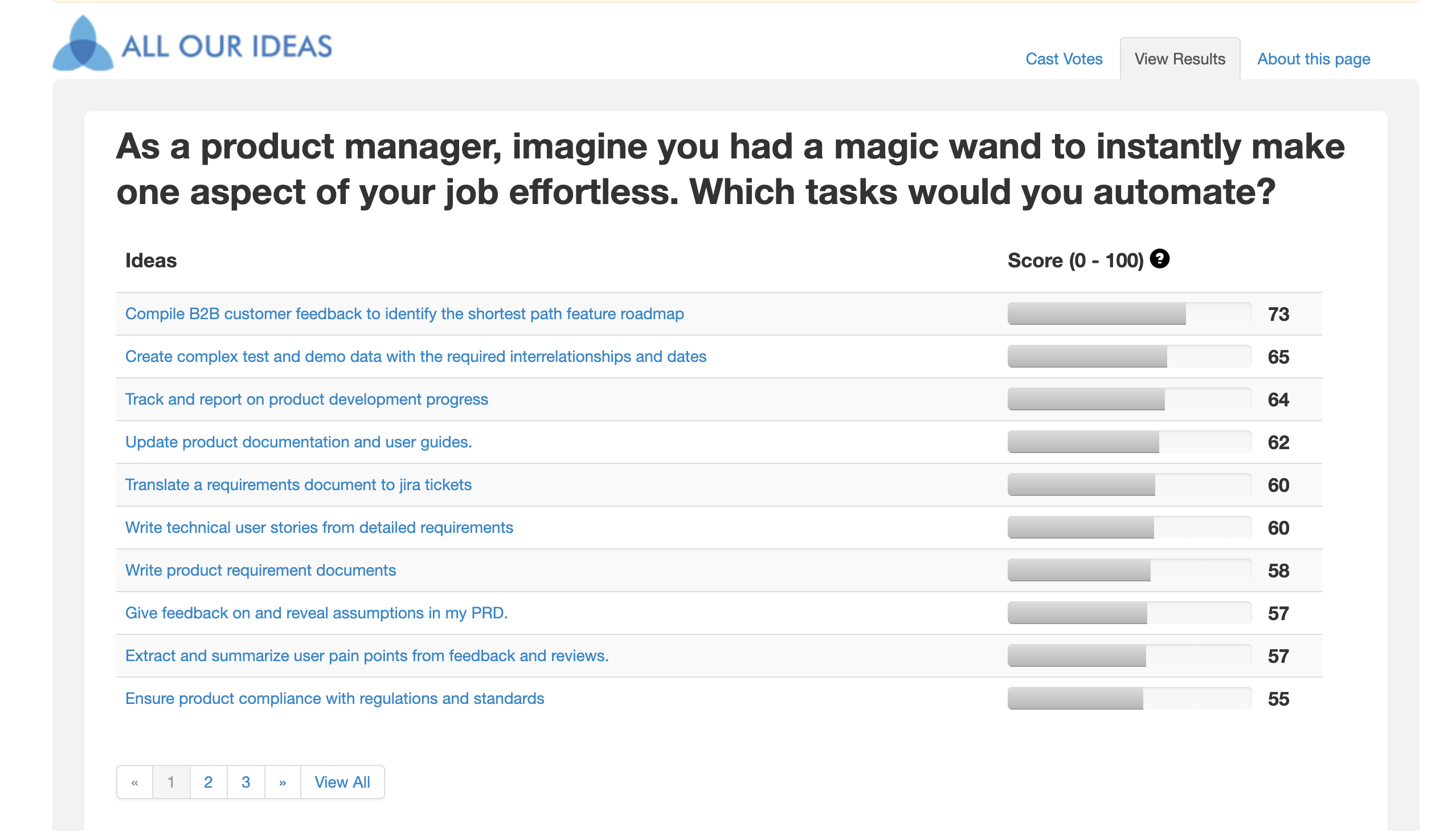}
        \caption{The top 10 results from the public poll, with each idea’s score representing its estimated likelihood of winning against a randomly selected idea (100 indicating a guaranteed win, 0 indicating a guaranteed loss).}
        \Description{The figure showcases the top 10 results of the public PM poll, with ``Compile B2B customer feedback to identify the shortest path feature roadmap'' being the top idea, and ``Ensure product compliance with regulations and standards'' being the 10th most popular idea.}
        \label{fig:magicwand2}
    \end{minipage}
\end{figure}

The design of human-AI interactions presents challenges due to uncertainty in the capabilities and the complexity of the output \cite{yang2020re}. Previous studies by Amershi et al. \cite{amershi2019guidelines} and Weisz et al. \cite{weisz2024design} used modified heuristic evaluations to create design principles for AI tools, asking participants to choose AI interfaces/tools and evaluate them for applications and violations of their own design guidelines. Building on this approach, we created Yodeai to provide a hands-on experience, allowing us to observe and investigate the interactions of knowledge workers with AI features in real time \cite{li2024user}.

Our design process began with a basic chatbot interface, building on PMs' familiarity with AI tools. To validate and expand our formative study findings, we conducted a public poll on All Our Ideas with 111 PMs found on LinkedIn (Figure \ref{fig:magicwand1}), asking them to rank and propose features that could improve their PM workflow. The poll surfaced additional use cases like generating Jira tickets and product requirement documents (Figure \ref{fig:magicwand2}), complementing our formative study insights. While poll respondents showed interest in product document generation and competitive analysis tools, our formative study participants emphasized a more fundamental need for systemizing user research processes. In particular, the poll highlighted both “compile B2B customer feedback” and
“extract and summarize user pain points from feedback and reviews”, while participants like E12 wanted tools for theme identification, and E20 expressed interest in a tool for organizing customer research and interrogating
data through chat. E10's observation about the lack of a unified platform for information reuse and revisit led to our development of reusable, shareable, and customizable widgets. We focused on interview and review analysis tasks as they mirror common PM activities \cite{schirr2013user} and parallel similar tasks in other knowledge work domains like consulting and academia, making our findings more broadly applicable.

Thus, we decided to focus on three specific widgets (i.e., Q\&A, User Insights, and Pain Point) inspired by the poll and the formative study.

\subsection{Organization \& Structure, Context \& Integrations: Overall System}
\begin{figure}[h!]
\centering
\includegraphics[width=\linewidth]{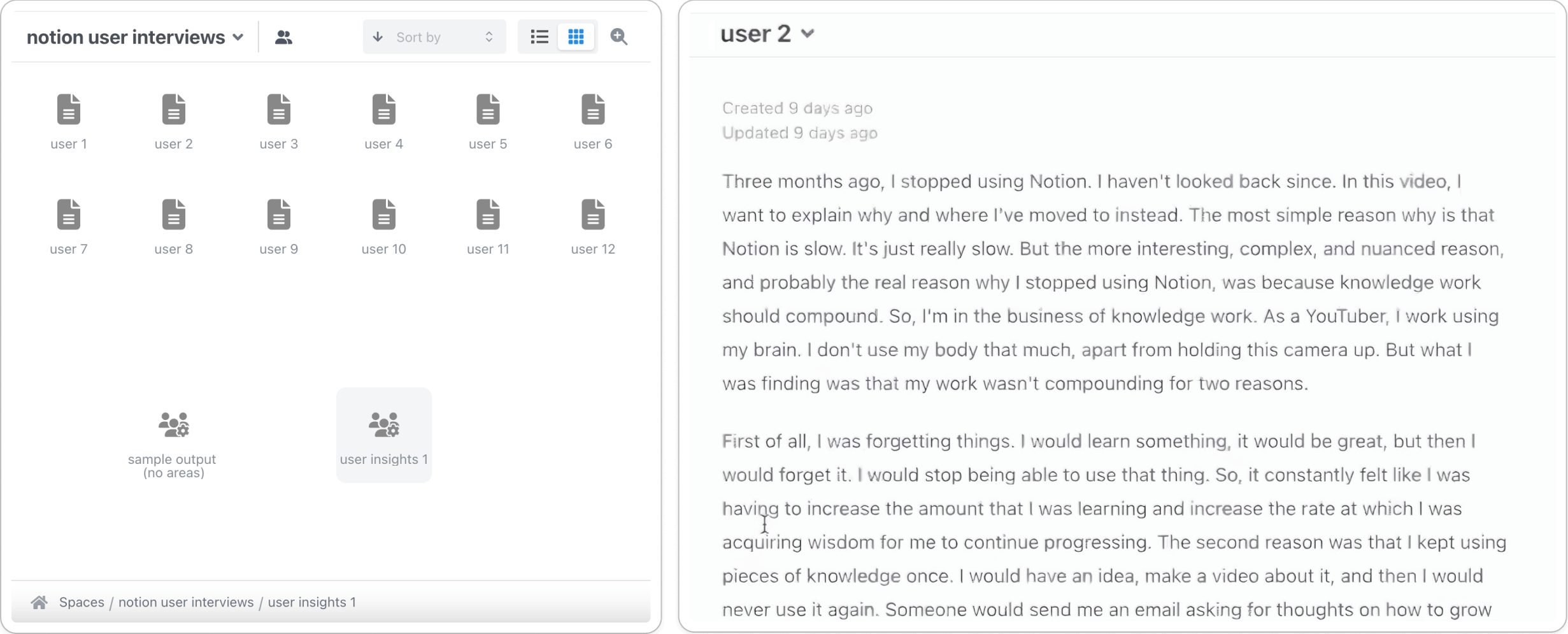}
\caption{Yodeai's space and page structure. Left: A space containing multiple pages including raw data (user interviews) and widget outputs (user insights). Right: Individual page view showing an interview transcript. This organization allows users to maintain connections between source data and derived analyses.}
\Description{The figure shows two side-by-side views of Yodeai's interface. The left panel shows a space containing multiple pages, including individual user interviews labeled ``User 1'' through ``User 10'' and widget outputs like ``User Insights 1''. The right panel displays the contents of a specific page ``User 2'', showing the full interview transcript.}
\label{fig:spaces}
\end{figure}

Based on the formative study's emphasis on having a centralized platform that carries context, Yodeai offers a structured system of pages and spaces (Figure \ref{fig:spaces}). A single space can contain multiple pages, including various data types such as text files, PDFs, spreadsheets, and whiteboards. Pages can be grouped flexibly across spaces and each of the pages have a generated AI summary of the content, giving users quick overviews with options for detailed exploration. Finally, the system facilitates collaboration through shared spaces and widget outputs.

\subsection{Prioritization \& Search: The Q\&A Widget}

\begin{figure}[h!]
 \centering
 \includegraphics[width=\linewidth]{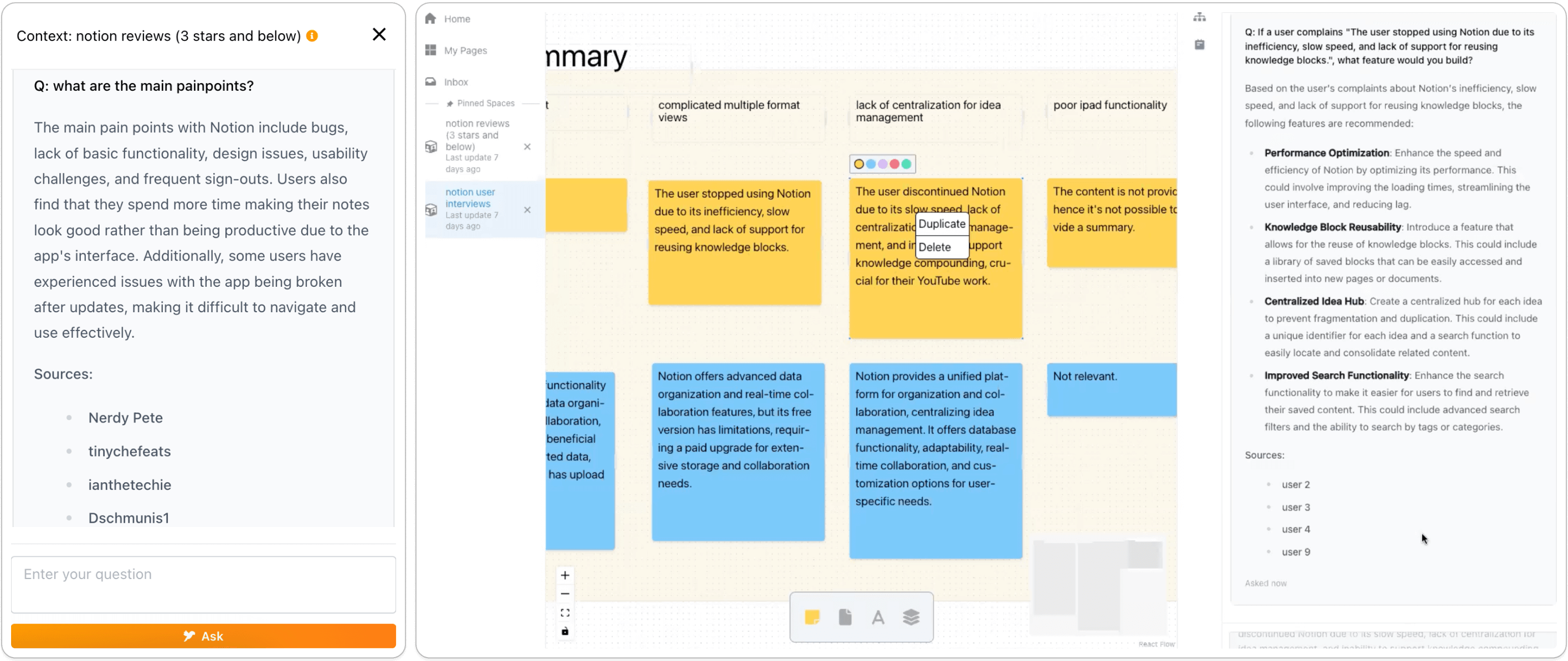}
 \caption{The Q\&A widget provides conversational access to data across different contexts. Left: Direct querying of notion reviews about pain points, with sources linked. Right: Analysis of User Insights widget output, where users can ask follow-up questions about patterns and themes identified in the sticky notes.}
   \Description{The figure shows two Q\&A interactions side by side. The left image shows Yodeai's Q\&A widget analyzing notion reviews, with the context labeled as ``notion reviews (3 stars and below)'', where someone asks ``what are the main painpoints?'' with the answer and source links displayed below. The right image shows the Q\&A widget being used to analyze a User Insights widget output, where users can ask questions about patterns identified in the sticky notes and get synthesized responses.}
   \label{fig:qa}
\end{figure}

The Q\&A widget (Figure \ref{fig:qa}) operates on the data in the current active space, allowing
team members to ``talk to the data'' and make sense of new information repositories, as needed by the formative study participants. This promotes efficient
collaboration and reduces the time spent on manual data organization and communication that often comes
with sharing data \cite{smirnov2023towards, crescenzi2019towards}. This feature promotes increased engagement and control \cite{marchionini2006exploratory} and enables intuitive navigation. For example, if a user is working in a space containing multiple user interviews, they can ask the Q\&A widget to identify users who mentioned a specific topic. The widget will then retrieve the relevant
information along with sources for easy cross-referencing, as requested by the formative study participants.

\subsection{Organization \& Structure: The User Insights Widget}

\begin{figure}[h!]
    \includegraphics[width=\linewidth]{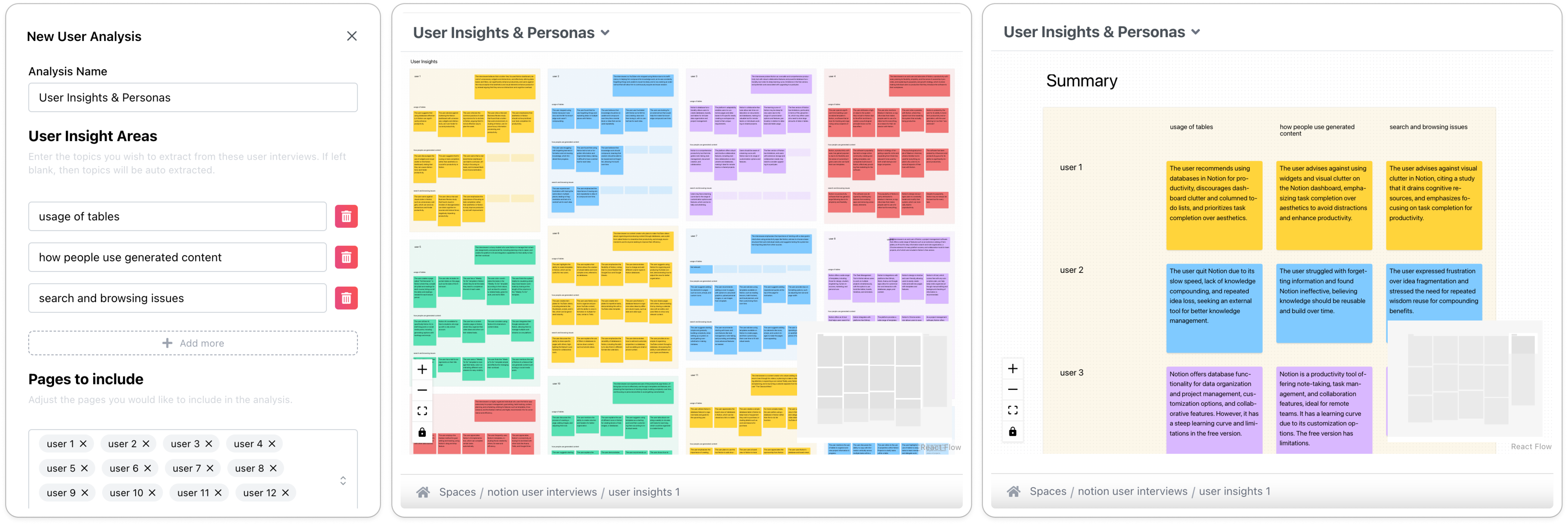}
    \caption{The User Insights Widget allows users to input insight areas and select pages for analysis. The widget generates a whiteboard view with sticky notes presenting user-specific insights and a summary, grouped into relevant areas of analysis.}
    \Description{Yodeai's User Insights Widget, represented by 3 images. The first image showcases the input box/modal for the widget, with the title of the modal being ``New User Analysis'', and 3 inputs for User Insight Areas: usage of tables, how people use generated content, and search and browsing issues, as well as Pages to Include and a multi select option to pick which pages to analyze. The second image showcases the whiteboard output, with 3 rows of 4 users and their associated sticky notes. The last image showcases the summary (also included in the whiteboard output) of what the first 3 users said about each topic.}
    \label{fig:user_insights}
\end{figure}

The User Insights widget (Figure \ref{fig:user_insights}) transforms unstructured user data into an interactive whiteboard visualization—a design choice that emerged from our formative study findings about PMs' need to synthesize overall patterns. The widget automatically clusters user data in a space (e.g., interview transcripts) and presents dual-granularity summaries \cite{simon1990bounded}: lower-level sticky notes showing each user's specific perspectives, and a higher-level synthesis of patterns across all documents. Users can specify multiple areas of analysis to focus on or leave it blank for full automation, allowing them to understand themes and sentiments without reading lengthy text, while still maintaining some control \cite{kim2023cells}. This flexibility emerged from our observations of how some PMs want a system to automatically identify themes for them.

\subsection{Organization \& Structure, Context \& Integrations: The Pain Point Tracker}
\begin{figure}[h!]
   \includegraphics[width=\linewidth]{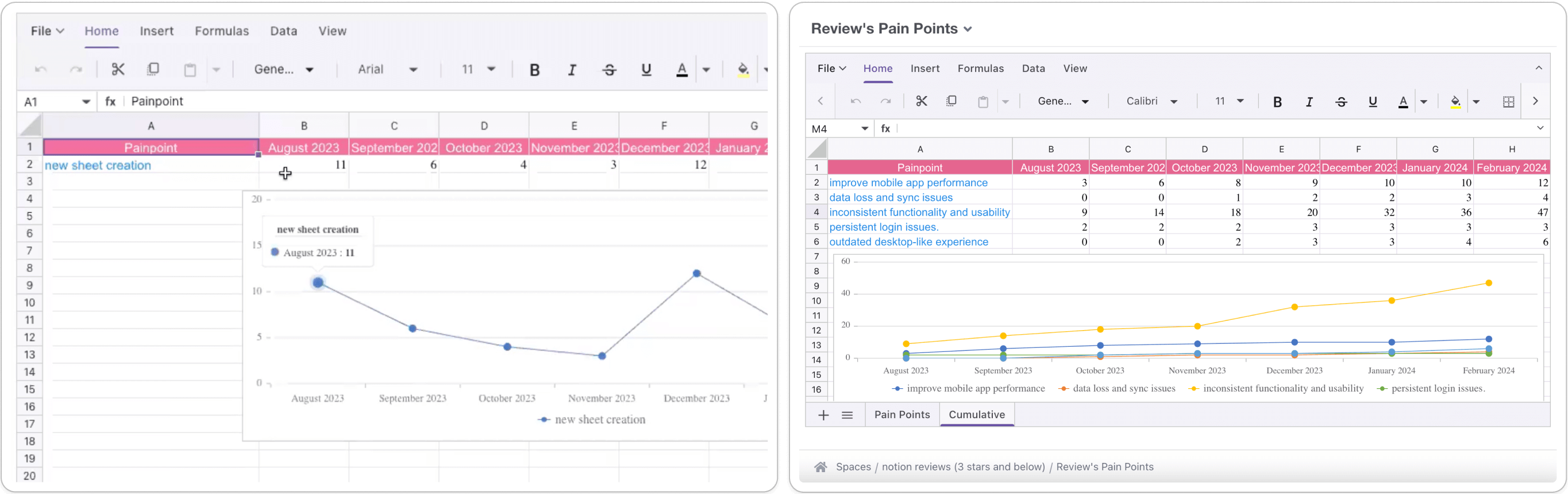}
   \caption{The Pain Point Tracker's output varies based on user input. Left: Auto-generated analysis identifying and tracking general pain points over time. Right: A more focused analysis based on user-specified areas of interest, showing how providing additional context to the LLM yields different insights. Both views include spreadsheet formats and cumulative graphs to support temporal analysis.}
   \Description{The figure shows two Pain Point Tracker outputs side by side. The left shows an auto-generated analysis with basic pain point categories and their frequency over time. The right shows a more nuanced analysis where the user has specified particular areas of interest, resulting in more targeted pain point categories and detailed temporal patterns. Both include spreadsheet views of monthly counts and cumulative graphs showing trends over time.}
   \label{fig:pain_point}
\end{figure}

The Pain Point Tracker (Figure \ref{fig:pain_point}) enables iterative quantitative analysis of user feedback through varying levels of user guidance, as requested by participants in the formative study who worked with quantitative data. When run in auto-generation mode, it independently clusters data to identify main pain points, while in guided mode, it incorporates user-specified categories, reducing the need for prompt engineering \cite{zamfirescu2023johnny}. This flexibility allows PMs to explore data from different angles: starting with automated clustering to discover unexpected patterns, then iteratively refining the analysis by providing more specific guidance based on their domain knowledge or hypotheses. The widget consistently provides both spreadsheet views of occurrence counts and cumulative graphs over time, supporting prioritization decisions based on observed patterns and perceived impact \cite{kahneman2013prospect}. Finally, since the formative study revealed the need for integrations with outside platforms, the Pain Point Tracker has an option to first pull data from Apple store reviews, considering the rating and date metadata, to populate the space, before running its analysis.

\subsection{Widget Flexibility and Adaptability}
\begin{table*}[h!]
\centering
\caption{Overview of Widget Modalities, Data Types, and Use Cases}
\includegraphics[width=\linewidth]{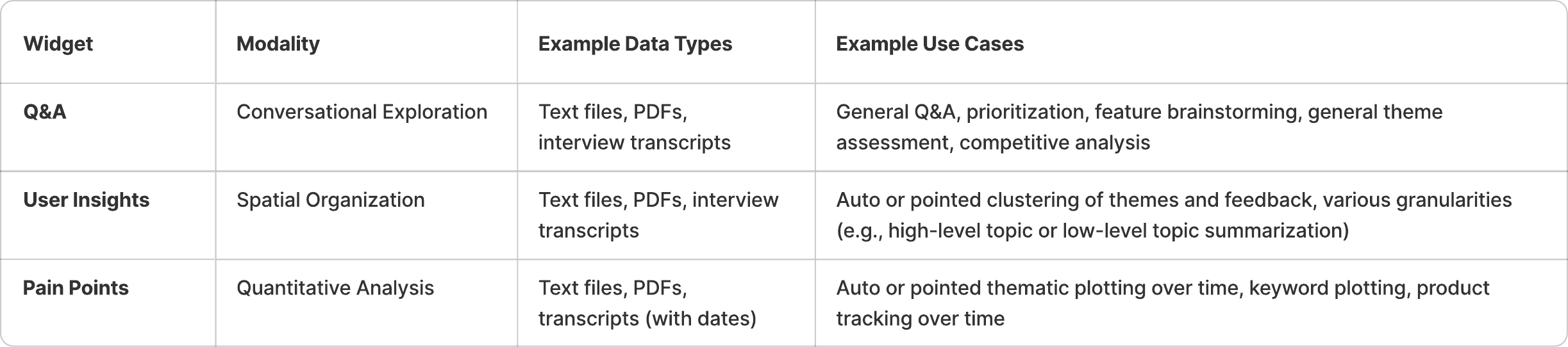}
\Description{Table summarizing the three distinct widgets: Q\&A, User Insights, and Pain Points. Each widget is categorized by its modality (Conversational Exploration, Spatial Organization, Quantitative Analysis), the types of data it uses (e.g., text files, transcripts, user reviews), and its typical use cases (e.g., general Q\&A, clustering, thematic plotting).}
\label{tab:widgets}
\end{table*}

\begin{figure}[h!]
   \includegraphics[width=\linewidth]{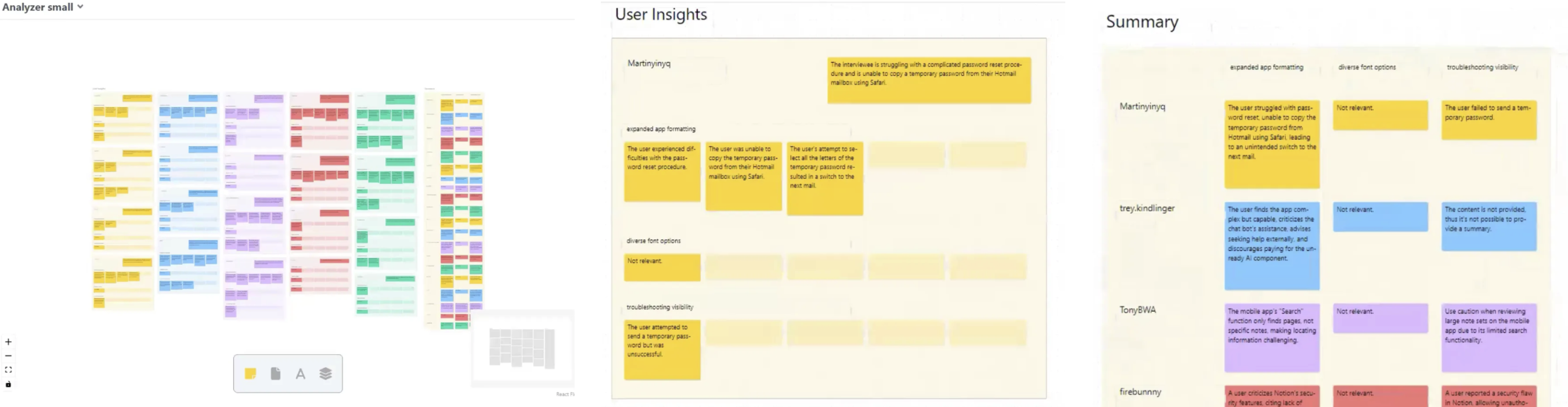}
   \caption{The User Insights Widget used on Apple Store reviews.}
   \Description{The figure shows a more crowded user insights widget output, with much of the text saying ``Not relevant''. }
   \label{fig:user_insights_wrong}
\end{figure}

\begin{figure}[h!]
   \includegraphics[width=0.7\linewidth]{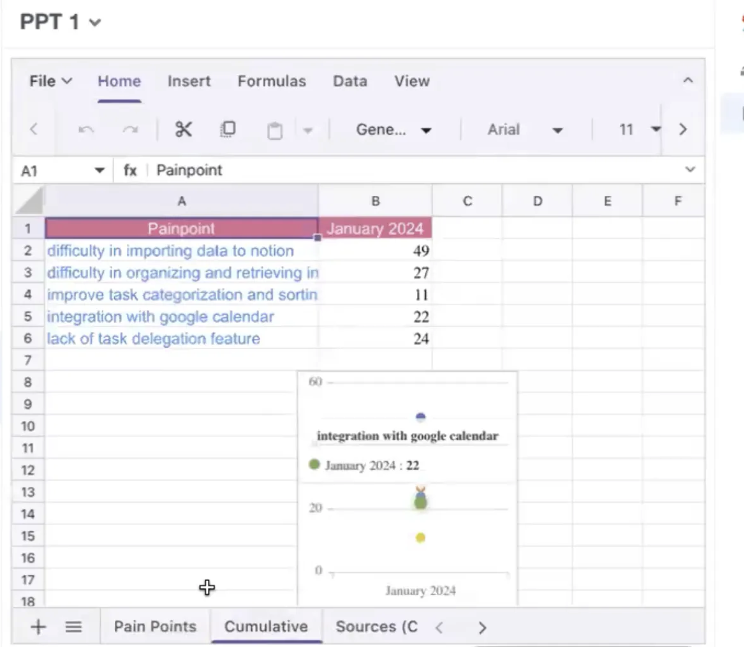}
   \caption{The Pain Point Tracker used User Interviews with no date.}
   \Description{The figure shows two Pain Point Tracker outputs after it has been used on no date user interview transcripts, resulting in all the data points being graphed on top of one another. The painpoints such as ``difficulty in importing data to notion'' are still listed in the graph.}
   \label{fig:pain_point_wrong}
\end{figure}

The flexibility to create multiple instances of each widget, with varying levels of user control from fully automated to user-directed, encourages experimentation and iteration \cite{kim2023cells, weisz2023toward, weisz2024design}. For example, PMs can create multiple User Insights widgets to analyze the same interviews through different thematic lenses, or several Pain Point Trackers to compare issue trends across different user reviews, supporting both divergent thinking (exploring multiple perspectives) and convergent thinking (synthesizing findings).

Furthermore, the widgets have minimal requirements regarding the types of data they can process (Table \ref{tab:widgets}). The Q\&A widget breaks all the pages in a space into chunks and uses both the pages' content and previous history as context when answering questions. The User Analysis widget assumes that each page is a user interview or an entity that the PM wants to separately analyze. Thus, the User Analysis widget can even be used on Apple store reviews, although the output would not be nearly as detailed as for interviews (Figure \ref{fig:user_insights_wrong}). Similarly, the Pain Point Tracker assumes the data to have timestamps to be able to graph across time; if used on user interviews without timestamps, the pain points it identifies will still be useful, though the graph would not (Figure \ref{fig:pain_point_wrong}).

\subsection{Implementation Details}

Yodeai is a web application built using Next.js for the frontend, Supabase for data storage and vector search, and Python for the backend. The widgets are powered by GPT-4, a large language model developed by OpenAI. For the Q\&A widget, we use retrieval augmented generation (RAG) and obtain the top 5 chunks related to the user question with vector search, feeding it as context to GPT-4. For the User Insights widget auto-generated topics, we use vector averaging to get the top 5 chunks that are related to a single page, and generate three main topics, iteratively updating them with each page. In terms of generating the content for each user and topic, we also use RAG to retrieve relevant portions of the interviews, summarizing them with GPT-4. Finally, the Pain Point Tracker uses k-means clustering to cluster evenly segmented chunks of all the reviews, and then retrieves the 5 most representative chunks of each cluster to serve as context for GPT-4 to determine an overarching pain point. We use additional hyperparameters such as cosine similarity thresholds or relevance scores generated by GPT-4 to exclude chunks that are too far from the center, and then proceed to total the count of reviews for each pain point. For embedding words into vectors, we use the bge-large-en model hosted on Hugging Face. Finally, we use React Flow for the whiteboard integration, SyncFusion for the spreadsheet, and Mantine for UI components. All code is open source: https://github.com/yodeai/yodeai.
\section{User Study}

With Yodeai, our aim was to answer the following question: \textit{What opportunities and limitations does generative AI pose for data navigation and decision-making in product management?}

\subsection{Participants}

We recruited 16 PMs in the United States (age: M = 31.25, SD = 9.46; gender: 5F, 11M) through a survey shared on LinkedIn. 15 of them identified as PMs and 1 as a developer advocate. All of them had several years of experience (M = 6.31, SD = 6.97), currently work in companies ranging from startups to large corporations, and explained that their day-to-day work involves a significant amount of PM tasks. Every participant mentioned that synthesizing user insights and tracking and evaluating pain points are within their primary responsibilities. 14 participants were familiar with AI tools, stating that they use an LLM (ChatGPT, Gemini, or Claude) in their work, ranging from daily to every two weeks. Every participant was compensated with a \$50 Amazon gift card.

\subsection{Procedure}

Each semi-structured interview lasted approximately 80 minutes and was divided into five main sections. The interviews were conducted remotely over Zoom, and participants were asked to share their screens during the system onboarding and main task sections. Each interview was conducted by two interviewers (both authors), who took turns conducting the pre-/post-interview questions and the task itself while the other person took notes. All procedures were approved by the IRB of our institution and informed consent was obtained from each participant.

\subsubsection{Pre-Task Interview (15 minutes)} 

The pre-task interview aimed to gather information about the participants' professional background and their current practices related to product management, user/pain point analysis, and AI usage. 

\subsubsection{System Onboarding (20 minutes)} 

During the onboarding process, we shared with participants Yodeai spaces containing publicly available data about Notion, an app for note taking and information organization. Specifically, we had two spaces: an Interview space, which contained 12 long and unstructured YouTube transcripts, and a Review space, which contained 132 reviews from the Apple App Store that were 3 stars and below. We started by explaining to participants how to use the Q\&A widget, prompting them to ask a question related to the data in the spaces, such as ``give me a pain point users have when using Notion,'' to showcase how the system pulls information from the spaces containing interview transcripts and scrapes relevant data to provide a response. Next, we showed them how to use the Pain Point Tracker by auto-generating a couple of pain points from all the reviews. Finally, we introduced the user insights widget by having participants generate a whiteboard view of a subset of the interviews.

\subsubsection{Task (25 minutes)}

The main task was a role-play scenario in which participants assumed the role of a PM at Notion. Using the two spaces containing all relevant information, participants were instructed to use Yodeai to propose 6 new feature or bug fixes (divergent phase), and then prioritize the top 3 proposals (convergent phase). During the task, participants were encouraged to articulate their thought process and concerns as they interacted with Yodeai.

\subsubsection{Post-Task Interview (20 minutes)}

Following the main task, we conducted a semi-structured post-task interview to gather participants' feedback on their experience using Yodeai, as well as general comparisons with existing tools. We asked open-ended questions to encourage participants to share their thoughts and reflections on the tool and the use of AI in their work.

\subsubsection{Post-Interview Survey}

After the interview, we asked participants to complete a survey to collect quantitative data on various aspects of the study, including the task design, participants' experience with Yodeai, and their data exploration and decision-making processes. 


\subsubsection{Data Analysis}

\begin{figure*}[h!]
\centering
\includegraphics[width=0.7\linewidth]{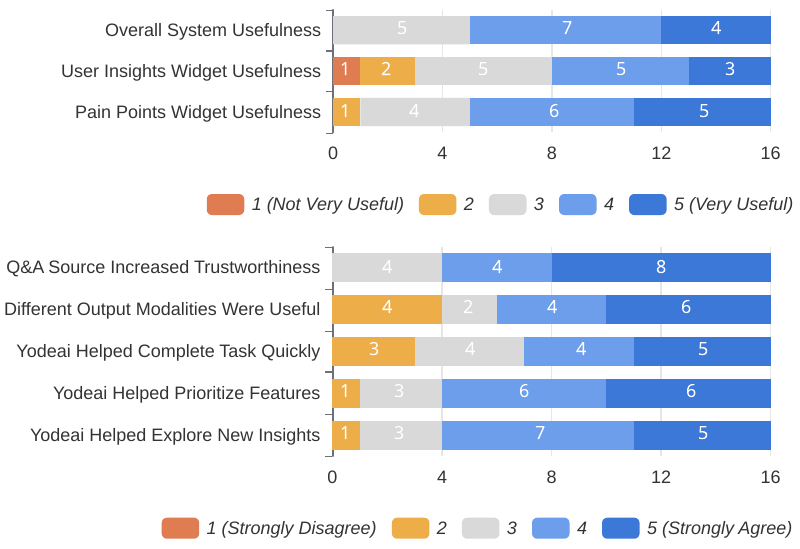}
\caption{Evaluation of Yodeai: most participants found Yodeai's widgets useful for exploring and prioritizing features.}
\Description{The results of an evaluation of Yodeai, with two bar charts. The first bar chart has 1 to stand for not very useful, and 5 for very useful. For overall system usefulness, 5 participants rated a 3, 7 rated it a 4, 4 rated it a 5. For user insights widget usefulness, 1 rated it a 1, 2 rated it a 2, 5 rated it a 3, 5 rated it a 4, 3 rated it a 5. For Pain Point Tracker usefulness, 1 rated it a 2, 4 rated it a 3, 6 rated it a 4, 5 rated it a 5. The second bar chart has 1 for strongly disagree, and 5 for strongly agree. For Q\&A source increased trustworthiness, 4 rated it a 3, 4 rated it a 4, 8 rated it a 5. For different output modalities were useful, 4 rated it a 2, 2 rated it a 3, 4 rated it a 4, 6 rated it a 5. For Yodeai helped complete task quickly, 3 rated it a 2, 4 rated it a 3, 4 rated it a 4, t rated it a 5. For Yodeai helped prioritize features, 1 rated it a 2, 3 rated it a 3, 6 rated it a 4, 6 rated it a 5. For Yodeai helped explore new insights, 1 rated it a 2, 3 rated it a 3, 7 rated it a 4, 5 rated it a 5.}
\label{fig:eval}
\end{figure*}

We analyzed the interviews using thematic analysis \cite{braun2012thematic}, with at least two authors coding each interview to ensure reliability. We employed line-by-line coding for both pre/post-task and task sections, creating inductive codes (e.g., ``Trust and Explainability'', ``Summarization''). This bottom-up approach \cite{bingham2021deductive} was more suitable for our semi-structured interviews than a top-down method, as the high level research questions were too broad to effectively group the findings, and reading through the transcripts and notes from the interviews line by line allowed for detailed grouping and labeling that could eventually be clustered together, to ensure that as much insight was gleaned as possible. We then used affinity diagramming \cite{harboe2015real} in FigJam to cluster quotes and actions, identifying themes and sub-themes. Finally, we selected relevant sub-themes for data navigation and decision-making, ensuring a comprehensive analysis aligned with best practices in HCI qualitative studies. We use prior work to help interpret and generalize our findings, which are placed on the backdrop of how PMs utilize AI tools.

\section{Results}

\subsection{What opportunities and limitations does generative AI provide for data navigation?}

\subsubsection{\textbf{Opportunity \#1: Providing a starting point at differing levels of abstraction}}

\paragraph {Yodeai Specific Findings:}

During the exploration and research stage of the user study task, participants (P3-P6, P16) appreciated how Yodeai helped them navigate large amounts of data and condense raw information into different levels of abstraction. Participants (P1, P3-P4, P6, P10, P13) highlighted the efficiency gained in navigating and making sense of new information using the proxy system Yodeai; they saw the widgets as starting points for their research process. P3 mentioned that Yodeai offers a ``first pass of what the data looks like,'' while P15 noted that the outputs help them see what people are mostly mentioning to get a ``general sense of what people like about the product.'' P5 also noted that Yodeai is particularly helpful when ``dealing with a high volume of external data that you are not directly familiar with.''

Participants explained that creating whiteboards with sticky notes or spreadsheets of pain points are familiar tasks in their day-to-day jobs, and without the generative AI widgets in Yodeai, the entire process would have been manual. Each participant had a different preference for the outputs of the widgets, as shown by Figure \ref{fig:eval}. Some participants, such as P13, found the granularity of the Pain Point Tracker too general and used the Q\&A widget to narrow down the pain points, while others, such as P16, appreciated the high-level categorizations. Participants like P6 liked the in-depth breakdown for each participant in the User Insights widget whiteboard, while others, such as P8, gravitated towards the high-level summary, finding the sticky notes overwhelming. The 16 participants stated that the Q\&A widget was a crucial part of their workflow, the sources allowing them to navigate and find specific pages without reading each one. For example, P4 was able to find two user interviews that specifically mentioned ``overwhelm of complexity'' by asking for references.

\paragraph {General Findings:}
Many PMs echoed the notion that AI helps them combat ``blank page syndrome'' and gives them starting point content and ideas (P1, P7, P13). For example, P13 stated that if they wanted to learn more about enterprise collaboration, they would use ChatGPT to ask questions such as ``What are the different ways of collaborating via commenting or message threads?''
Additionally, PMs deal with high variability in the type, quantity, and granularity of data. P16 noted the difference between building version 1 and version 10 of a product; with version 10, there is a lot of user data and logs to sift through, while for version 1, the data would primarily come from high level competitive analysis, as used by P5, P8 and P12. P15 also clarified the different types of data they have to work with: quantitative usage analytics vs qualitative user interviews, while P8 disclosed that they use surveys, user interviews, and comparative analysis. In addition, when presenting findings to various stakeholders, the presentation format differs. P1 noted that they would use a whiteboard-like output to present to the CEO, while P9 writes detailed PRDs (product requirements document) with a list of interviews and competitors for their engineering team. Finally, the level of seniority also influences the granularity of work that a PM does; P13, a senior PM, stated ``I'm more involved in the high level strategy and the high level prioritization of features, whereas a more junior PM [...] would still do a cycle of going even deeper into the customer needs.''

\subsubsection{\textbf{Opportunity \#2: Organized and transparent collaboration through an audit trail}}

\paragraph {Yodeai Specific Findings:}

Most participants (12/16) appreciated Yodeai's shareable spaces and pages, agreeing that its structure facilitated transparent collaboration. Nearly all (15/16) found sharing widget outputs and Q\&A results beneficial for teamwork. P8, while acknowledging Yodeai's helpfulness for shared information, felt the outputs were ``too raw'' for direct team presentation. P16 specified they would share within the product team but not with marketing or sales, highlighting the need for polished results for certain team members. Yodeai's raw data collection allows for exploration by team members involved in data navigation. P12 noted its particular benefit for larger product teams, enabling independent exploration of interviews and data. P7 and P9 emphasized the importance of visibility and transparency regarding feature status and data for stakeholders. P9 and P10 stressed the need for engineers to develop empathy, with P9 stating, ``if I share [Yodeai] with engineers, it's gonna be useful in that they would get a firsthand view of what users are actually talking about [...] Engineers never talk with users. And it's the PM, and the salespeople who always talk with the users, so it's our duty to share those insights with the engineers, so that they are also empowered by the problem they're solving.''

\paragraph {General Findings:} 

Participants noted collaboration within and outside their immediate teams. P10 mentioned working with marketing, designers, engineers, customers, and partners, while P2 emphasized collaborating with VP, CTOs, and developers. Each cross-functional partner requires different outputs and data synthesis. P7 explained, ``depending on the use case, customer success managers want to know the specifics of when a feature will come out so they can tell the customer. And for engineers and designers, it would be more about the creativity of creating different ideas as a collaborative effort.'' In addition, without an audit trail, P7 mentioned that sometimes people give a problem without context, so they have to spend two weeks to ``chase context''.

\subsubsection{\textbf{Limitation \#1: Overreliance and reduction of insight depth}}
\paragraph {Yodeai Specific Findings:}

Some participants expressed concerns about the quality of automated outputs. P2 felt a lack of connection with auto-generated sticky notes, stating, ``even though the data is tagged nicely on sticky notes, because these are auto-generated, I don't feel that connection with those sticky notes.'' In terms of the User Insights widget, P6 noted that the sticky notes were overly verbose when summarizing key points, preferring shorter, more focused content, while P7 felt ``blocked'' after seeing the summarizations and big themes, unsure of how to dig deeper.

\paragraph {General Findings:} 

Participants emphasized the importance of engaging directly with source data for deeper understanding. P4 explained that PMs ``cannot fully understand the issue without actually reading into the original text.'' While AI can improve the insights gathering process through summarizations and overviews, human intuition and contextual understanding generate more nuanced insights. P5 highlighted the value of direct customer feedback: ``Reading customers' direct words and getting the gist of it from AI are two different things. You understand more about the feelings and emotions behind the feedback when you look at the actual comments and words used by the customers.''

\subsubsection{\textbf{Limitation \#2: Outdated data and priorities}}

\paragraph {Yodeai Specific Findings:}

While participants appreciated Yodeai's centralization of information (P2, P10-P11, P16), some (P3, P8, P11) suggested improvements in navigation, such as the ability to overlay and combine information. P11 emphasized the importance of tracking pain points over time to identify outdated feedback. Many participants (P2, P11-P13, P15, P16) expressed interest in integrations from other sources like Reddit or LinkedIn, especially if these could auto-update with the latest information.

\paragraph {General Findings:}

Participants highlighted the challenge of maintaining up-to-date information when using LLMs. P5 noted that information can quickly become outdated, particularly with news updates. In terms of pain point identification, P11 stressed the need for ``constant revision'' in prioritizing pain points, emphasizing the importance of shifting focus as solutions are implemented. While LLMs help process large amounts of data, they struggle to identify stale or irrelevant insights over time; P2 stated that using AI would require constant extra context, like ``whether this pain point is already being worked on''. Users must continuously update the data fed into LLMs by adding new material and removing outdated information. P15 confirmed this, stating that his company either interfaces directly with customers or uses forums like Reddit to gather current feedback, while P7 uses customer success conversations or quarterly surveys.
\subsubsection{\textbf{Limitation \#3: Privacy and confidentiality}}

\paragraph {Yodeai Specific Findings:} 

Participants expressed hesitancy about using Yodeai professionally without company approval. P12 emphasized the need to ensure ``confidentiality of my company's source content.'' Participants were more likely to trust tools with integrations from well-established companies like Slack or OpenAI (P2, P9), with P13 expressing delight that Yodeai had a Figma-like interface with the User Insights widget.

\paragraph {General Findings:}

Companies are cautious about adopting external AI tools due to confidentiality risks. Participants (P9, P11-P13) mentioned using only ``company approved'' tools. The rapidly evolving AI landscape complicates this, as approval processes may become outdated. As a result, many PMs use general-purpose tools like ChatGPT with redacted information (P3, P13) or rely on their company's enterprise chatbots (P6, P11-P12, P15).

\subsection{What opportunities and limitations does generative AI provide for decision-making?}

\subsubsection{\textbf{Opportunity \#1: Customization to diverse workflows}}
\begin{figure*}[h!]
    \includegraphics[width=\linewidth, height=5cm]{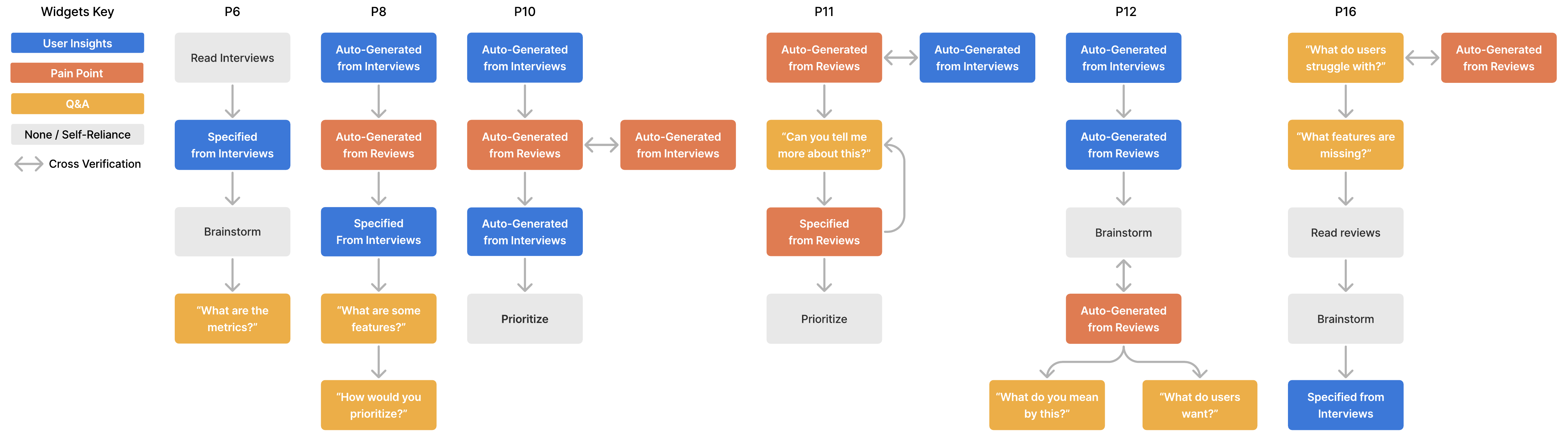}
    \caption{Visualization of 6 participants' decision-making workflows using Yodeai. The use of each widget at each step is indicated with a color: blue for User Insights widget, red for Pain Point Tracker, yellow for Q\&A widget with the question asked, and gray for no widget use. Double arrows represent cross-verification between steps. ``Interviews'' and ``Reviews'' refer to the respective Notion spaces in Yodeai, while ``Auto-Generated'' and ``Specified'' describe widget creation without or with pre-specified fields. This figure shows that each participant used the widgets in their own unique way, emphasizing the importance of tool flexibility.}
    \Description{The figure showcases 6 sample workflows by participants; for example, P6 first read interviews, then created a user insights widget with specified areas, then brainstormed on their own, then asked the Q\&A widget for metrics. Another example is P11, who first created a Pain Point Tracker with auto-generated reviews, cross checked it with a user insights widget with auto-generated areas, and then asked the Q\&A widget for more details about the Pain Point Tracker, then created another Pain Point Tracker with specified fields, and repeated the process of asking Q\&A widget for more details. After, P11 prioritizes on their own.}
    \label{fig:workflows}
\end{figure*}
\paragraph {Yodeai Specific Findings:}

Each of the 16 participants followed a unique workflow, demonstrating Yodeai's adaptability in assisting diverse decision-making processes. Figure \ref{fig:workflows} illustrates the variety in participants' approaches, including different starting points, ideation methods, cross-validation techniques, and prioritization strategies. Participants either combined content from both spaces into one feature list (P1, P2, P8, P12, P14, P16) or focused on one space (P6-P7). P5 highlighted Yodeai's flexibility: ``[Yodeai] is also open in a way because I can influence the results by providing specific keywords. So it's a mix of both. If I want to have control, I can, and Yodeai provides options for that. But if I don't want to, I can also just let the platform generate results based on its own analysis.'' All participants agreed on the usefulness of generating multiple widget outputs and editing focus areas. Some participants used widgets in unexpected ways, such as applying the Pain Point Tracker to user interviews or the User Insights widget to reviews (P2, P6, P8, P10, P12). In addition, some participants expected the different widgets to know about each others' outputs; for example, P4 thought that the Q\&A widget would have context about the graph generated to ask questions about it. Others wanted the Q\&A widget to know about the structure of Yodeai and meta information such as how many pages a space has (P9, P14).

\paragraph {General Findings:}
PMs highlighted the diversity of real-life workflows. For instance, P6 prefers manual
note-taking during interviews, while P9 records interviews for later analysis. PMs often juggle multiple roles, as noted
by P3 (UX researcher, data analyst) and P8 (design, engineering, legal issues). The level of data analysis also varies significantly between B2B and B2C companies. P16
explained: “[B2B] customers will tell you what they need and you don’t rely a lot on user research, but for B2C you
have customers that ask for features and you need to see the data and see how many people are using it and you need
to use other metrics to measure and decide.” P12 and P15 confirmed that B2B PMs focus primarily on accounts rather
than customer reviews and interviews.

\subsubsection{\textbf{Opportunity \#2: Intermixing of background knowledge}}
\paragraph {Yodeai Specific Findings:}

Participants combined personal heuristics with LLM reliance during exploration. P9 used product knowledge to guide their exploration, inputting specific features into the Pain Point Tracker. P7 leveraged the Q\&A widget and User Insight widgets for brainstorming. P6 mixed approaches, considering their ``own background and experience, and previous customer contacts or previous contacts with other products'' while reviewing Yodeai's summaries. P8 noted, ``even though I have some personal opinions on the user pain points based on a very brief review, once I throw that into the user insight widget, it could pretty effectively relate to whatever pain points users describe. That helps me make informed decisions regarding what features I'm trying to build.'' Some participants, such as P1, focused on self-specified topics and used personal heuristics to prioritize features like ``encryption'' and ``gamification of on-boarding.'' Others such as P11, P12, and P15 categorized bugs and features separately, emphasizing the importance of fixing bugs before implementing new features. This demonstrates how Yodeai's widgets help PMs validate and expand upon initial assumptions, leading to more informed decision-making.

\paragraph {General Findings:} 

PMs develop instincts and heuristics over time. P7 stated that senior PMs have experience to know ``I have tried this before in this context, and it didn't work out'' to quickly validate ideas. P15 noted that they are tightly intertwined with their product and have instincts on prioritization.

\subsubsection{\textbf{Opportunity \#3: Iterative prioritization}}

\paragraph {Yodeai Specific Findings:}

Participants employed a combination of Q\&A, personal experiences, and intuitions for prioritization. While some (P8, P14) utilized Q\&A during the convergent phase, others (P1, P5, P11, P15) relied solely on their experiences. The Pain Point Tracker provided a high-level overview, but participants often prioritized based on personal judgment. For instance, P12 categorized pain points into features or bugs, giving higher priority to bugs. Participants increased information granularity through various methods: they asked Q\&A for details (P1-P6, P11, P12, P14), referenced surrounding reviews and interviews (P2, P6, P16), and broke down points using new widget generations (P3-P4, P8, P11).  However, some participants wanted even more direction from Yodeai; P13 mentioned that they would want a list of next steps out of the insights, while P4 wanted to have effort estimation next to each of the next steps.

\paragraph {General Findings:}

P5 stated that in their previous PM experience, their manual process was to identify 5 most important pain points from customers, identify how the product would solve these points and what customers would gain, and then narrow it down to 2 pain points and build solutions. Thus, participants stressed the importance of quantitative data for decision-making (P6, P11, P14). P12 noted, ``as a PM, numerical data is hard to come by, any numbers is a big plus.'' Time-based information was valued for identifying stale feedback and tracking changing sentiments (P3-P5, P11-P12, P16). Some preferred displaying sentiment for each pain point (P9) or using visualizations like word clouds or pie charts (P9, P3).

\subsubsection{\textbf{Opportunity \#4: Facilitating data validation and collaborative trust}}

\paragraph {Yodeai Specific Findings:}

Participants emphasized the importance of cross-comparing and cross-verifying data for validating outputs. They appreciated the inclusion of sources in Q\&A outputs, which increased trust in LLM output (7 Strongly Agree, 5 Agree, 4 Neutral). Cross-validation methods varied among participants: some compared information between spaces (P3-P5, P10-P11, P16), others overlaid Pain Point Tracker graphs (P11), while some manually read source data (P2-P3, P6, P16) or used the Q\&A widget (P1-P3). Specifically, participants like P2 and P16 employed a mix of manual review reading, the Q\&A widget, and the Pain Point Tracker, leading to features such as ``create templates for task management based on domain'' (P2) and ``improve search functionality with highlighting'' (P16). Some participants, like P4, created both auto-generated and specified User Insight and Pain Point widgets to glean further details on what features to build. Participants valued Yodeai's audit trail feature for team collaboration and decision-making transparency. P9 stated they would share spaces with managers and mentors, noting that ``managers are very much interested in how you are identifying the problem because the problem is important.''

\paragraph {General Findings:}

For personal work with generative AI, PMs adopted various approaches to mitigate misinformation. Some avoided ChatGPT for research due to hallucination fears (P9), while others used AI only for summarization or revisions (P3, P9). Cross-verification methods included checking with Google (P1, P7), requesting sources from AI (P2, P12), and comparing against company documentation (P5, P8). P11 noted that ``a lot of PMs are not willing to risk their job on the output of the model,'' emphasizing the need for constant data efficacy checks. Many of the PMs echoed that their process for pain point and user interview analysis is manual due to the need for validation. They stated that their first step to user interview analysis was to read over their notes or the transcripts, as many of them wanted to be familiar with the raw data before starting out (P1, P5-P6, P8, P12, P14, P16). In terms of the pain point identification process, P9 and P16 stated that they would read through the actual reviews word for word. When they did use AI, many of the PMs used it as a direct Q\&A format to ask questions about the interviews (P10, P12), emphasizing that they would be able to ``check'' the AI since they were already familar with the raw data (P4). However, they noted that when they did use chatbots such as ChatGPT, they would have to keep reprompting and asking questions to structure the answers better (P1, P3, P7).

Due to the collaborative nature of PM work, PMs echoed that it is also difficult to verify the work of others;  P11 shared that one of their tasks as a PM was to blindly trust an Excel sheet of pain points from users without knowing the data’s origin or how it was analyzed.

\subsubsection{\textbf{Opportunity \#5: Trade-off between creativity and objectivity}}

\paragraph {Yodeai Specific Findings:}

During the identification of new bug fixes and features, PMs utilized both objective information from the widgets and data, and was creative either by themselves or with Yodeai. For example, P6 ideated features like ``30 days trial version of paid premium subscription'' and ``nudges on old tasks/notes'' based on the auto-generated User Insights widget output and their past experience, while P5 prioritized high-level bugs like ``login and access issues'' and ``data loss'' using the auto-generated Pain Point Tracker and their personal intuitions. Others used AI for creativity, such as P8, who heavily used the Q\&A widget to brainstorm and prioritize features, resulting in a final list that included features such as ``support for Apple pencil for handwritten notes and sketches.''

\paragraph {General Findings:}

Participants emphasized the value of leveraging LLMs' creative aspects in decision-making. PMs such as P7 would prompt engineer by giving ChatGPT a pain point, provide context as a PM and their specific company, and ask for suggestions. P1 also highlighted ChatGPT's potential in the ideation phase, and the fact that regenerating is useful to avoid fixation: ``The answers [ChatGPT] gives can spark new ideas in the ideation part.'' P10 also discussed how LLMs can uncover insights by combining data from disparate sources and handling edge cases and knowledge gaps. In such scenarios, prompting may not be very useful, as users need to recognize that ``new insights could be something that we were not thinking about'' and may not be present in the existing dataset. However, PMs such as P9 and P11 echoed the notion that they used AI to summarize or edit information rather than spout new insights. For instance, P5 leverages LLMs for background research and writing assistance due to English not being their first language, whereas P14 uses ChatGPT to refine information they are already familiar with.

\subsubsection{\textbf{Limitation \#1: Contextual factors outside of the model’s reach}} 
\paragraph {Yodeai Specific Findings:}

Participants emphasized the importance of considering factors beyond Yodeai's outputs when making critical decisions. These factors include business context, collaborator opinions, monetization benefits, and product strategy (P4-P6, P11, P12, P15, P16). P6 noted, ``at the end of the day you still need a person who actually knows the business context to look at and review the feedback given by the tools.'' Additionally, participants (P11, P13, P16) stressed the need for an export feature in Yodeai to work on widget outputs externally, such as exporting to Figma for design teams or overlaying Pain Point Tracker outputs with their product roadmap to see what happened when they shipped.

\paragraph {General Findings:}

Participants highlighted the high-stakes nature of their work environment, and noted that they did not really use AI for prioritization, due to fear of hallucinations and other factors. P11 mentioned frequent data verification due to potential job loss from errors, while P2 noted the value of manual effort in ensuring correct decision-making. P11 emphasized the complexity of professional decision-making, including factors like budget, release schedules, marketing, and sales. P11 demonstrated prioritizing a pain point not listed as ``high priority'' by the Pain Point Tracker, emphasizing user retention: ``if an app crashes, a user is gone [...] that's why app crashing would be my first feature.'' P16 stressed the need to incorporate priorities from various stakeholders (sales, marketing, support teams, CEO) and continuously refine feature lists based on recurring issues, available resources, and company goals. P2, P6, and P15 stated that the job requires so much contextual knowledge and that it is not ``good enough'' to make decisions at work. P4 gave a bit of insight into the collaborative nature of the prioritization process, stating ``In my normal workflow [I] port [the list] to my engineers and see like the effort that's gonna require, and see if the effort corresponds to the impact that's gonna create again.''

\subsubsection{\textbf{Limitation \#2: Isolation and Confirmation Bias}}

\paragraph {Yodeai Specific Findings:}

P6 highlighted a potential limitation of Yodeai and similar AI tools, stating, ``it's good to be able to say you don't know the answer to something, or to literally be proven wrong, because that's how you learn. So if Yodeai is constantly telling me, 'Oh yeah, this is the right thing,' it could really lead me down the wrong path.''

\paragraph {General Findings:}

Participants reported using at least 24 different tools with AI integrations, highlighting the proliferation of AI in knowledge work contexts. While these tools offer potential benefits, they also raise ethical concerns. P7 expressed concerns about AI undermining collaboration: ``I don't know if AI actually enables collaboration because it's just like firing people and doing more with less. A single designer can come up with the ideas and not have anyone else involved, and collaboration kind of dies there.'' They also noted potential risks to creativity: ``I feel like AI is putting the whole creativity theme at risk [...] now you only need one person to create 30 ads instead of bringing people together to generate these ideas.'' P1 pointed out the tendency of AI, particularly ChatGPT, to be overly agreeable: ``I feel like often, to conform to my biases, it's just trying to be agreeable, I think, in general.''
\section{Discussion and Design Implications}

Our empirical study reveals three critical design implications for AI tools across knowledge-intensive and decision-making domains, initially examined through the lens of product management but broadly applicable to contexts requiring complex information synthesis and strategic decision-making. These insights emerge from studying how knowledge workers navigate data-rich environments where decisions carry significant organizational impact. 

First, AI tools must support flexible workflows that adapt to different work styles and analytical approaches. Second, they must enable robust verification and collaborative accountability given the high-stakes nature of professional decision-making. Third, they must balance the integration of diverse data sources with security concerns and evolving organizational priorities. We detail these implications in Table \ref{tab:AI_Design_Principles}, connecting each to PM specific tasks while highlighting their broader relevance to knowledge work.

\begin{table*}
\centering
\caption{Summary of Design Implications, Interactions, and Suggestions for Enhancing AI Tool Usability in Product Management and Knowledge Work.}
\includegraphics[width=0.95\linewidth]{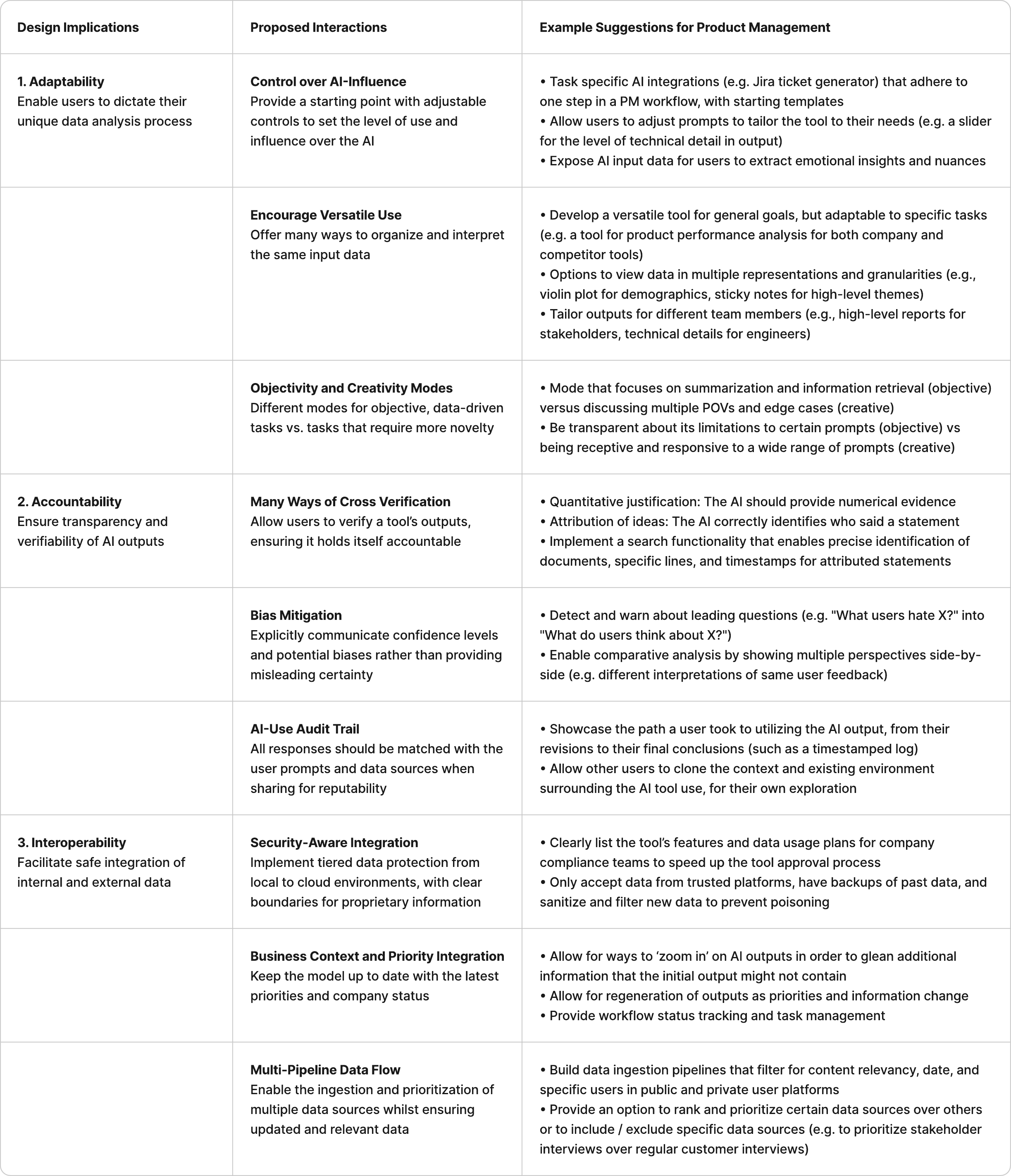}
\Description{A table summarizing design implications, proposed interactions, and suggestions for enhancing the usability of AI tools for knowledge workers. The design implications are: 1) Adaptability: Enable users to dictate their unique data analysis process, 2) Accountability: Ensure transparency and verifiability of AI outputs, and 3) Interoperability: Facilitate safe integration of internal and external data. For each implication, specific interactions surrounding control, versatile use, objectivity/creativity modes, cross verification, bias mitigation, audit trail, security integration, business context, and data flow are proposed. Concrete suggestions and examples are provided for putting these interactions into practice to create AI assistants that empower knowledge workers.}
\label{tab:AI_Design_Principles}
\end{table*}

\subsection{Design for Workflow Flexibility and User Autonomy}

\textit{PM specific task:} User Interview/Review Analysis

Rather than imposing new workflows, AI tools must adapt to how an individual PM goes through their tasks. Our study revealed that PMs have personal established and effective practices for analyzing user feedback. The role of AI should be to enhance these practices while letting PMs maintain control over how and when to use AI assistance.

\subsubsection{Control over AI-Influence} 

AI tools must let PMs dictate their preferred level of AI involvement rather than enforcing a fixed interaction model. Our study revealed stark differences in how PMs chose to work with AI: some relied heavily on automated features for efficiency (P8 completed tasks in 23 minutes using the Q\&A widget), while others deliberately limited AI use to preserve learning opportunities, preferring to read reviews and interviews manually to extract emotional context and engage in deeper reflection \cite{woodruff2023knowledge}. Even within individual workflows, PMs varied their AI usage—some used the User Insights widget solely for initial summarization before switching to manual analysis, while others leveraged the Pain Point widget throughout their process from identification to prioritization. This suggests AI tools should provide clear affordances for adjusting AI involvement at any point in the workflow, letting PMs fluidly shift between automated and manual approaches based on their current needs and preferences.

\subsubsection{Encourage Versatile Use} 

AI tools should be designed with the expectation of creative re-purposing rather than assuming fixed use cases. Prior work shows users tend to appropriate workflows for unintended tasks \cite{long2024not}, and our findings confirm this. For instance, P15 suggested using Yodeai to analyze competitor products when entering new markets, which was far beyond the tool’s original purpose, but showcases the versatility of the widget. In addition, different product lifecycle stages require different forms of data analysis, from early-stage competitive analysis to detailed feature refinement in mature products \cite{annacchino2003new}. Varying experience levels lead to diverse data representation preferences as well. The tool's combination of conversational, visual, and graphical formats proved valuable not because each format had a predetermined use, but because it helped PMs discover multiple entry points into their data; PMs often face large volumes of data and can easily become fixated on a single AI output, but diverse representations help reduce missed nuances, as shown by the PMs who tried to use different widgets on the same space. This adaptability was particularly crucial given the collaborative nature of PM work—the same analysis often needed to shift between detailed, unpolished data for product teams and high-level overviews for stakeholders. This also highlights the need for diverse data representations and options to export outputs; for example, Yodeai’s whiteboard results that emulate Figma enables collaboration with team members in other roles, like designers. Tools should therefore provide modular components (e.g. widgets) that can be repurposed rather than rigid, task-specific features. Additionally, AI tools should enable consideration of both input data and previous outputs, similar to how PMs wanted the Q\&A widget to incorporate insights from other widgets, though some, like P16, were concerned about the risk of self-pollution if global context was enabled.

\subsubsection{Objectivity and Creativity Modes} 

AI tools need to explicitly support and differentiate between objective analysis and creative exploration rather than conflating these distinct thinking modes. Our study revealed a clear tension in how PMs approached decision-making. Some preferred using AI exclusively for objective data consolidation while relying on personal judgment for creative insights, (such as P6 in Figure \ref{fig:workflows}) whereas others actively sought AI-generated suggestions for feature ideation (such as P8 in Figure \ref{fig:workflows}). This distinction reflects the dual nature of PM work: needing both rigorous analysis of user feedback and creative exploration of possible solutions. Tools should provide clear affordances for switching between these modes, letting PMs deliberately choose when to focus on objective data synthesis versus creative exploration. This separation helps PMs maintain clarity about when they're analyzing existing feedback versus exploring new possibilities, while still supporting natural transitions between these modes as they refine features and align them with business goals and engineering constraints.

\subsection{Ensuring Accountability and Validation in High-Stakes Decisions} 

\textit{PM specific task:} Feature and bug fix brainstorming

Product decisions have far-reaching implications for users, engineering resources, and business objectives. Our study revealed that PMs need comprehensive accountability mechanisms that go beyond simple AI verification—they need tools that help build and demonstrate decision confidence across multiple stakeholders.

\subsubsection{Many Ways of Cross Verification}

AI tools must support multiple complementary verification strategies that PMs can combine based on their context and confidence needs. Unlike personal AI use, where errors like hallucinations and misrepresentations may have
less impact, corporate AI use—especially in brainstorming product enhancements and fixes—directly affects company
success. Our findings reveal the need for robust cross-validation methods, such as source citation functionality that enables users to trace AI-generated insights to original data \cite{huang2023citation}. This was evidenced by how participants built confidence by combining multiple approaches: comparing findings across different data spaces, generating multiple widget outputs to compare with each other, and frequently returning to the source data. Thus, AI tools should allow direct comparison of outputs at various levels, such
as comparing regenerations of the same or different AI tools, comparing outputs with the same tool and parameters
but different data, or comparing outputs from different tools with the same data. This not only allows PMs to be able to see common themes brought up by several different tool outputs, but also allows them to see the outlier outputs
that they might have to investigate further.

\subsubsection{Bias Mitigation}

Tools must help PMs identify and counteract potential biases in both AI suggestions and their own decision-making processes. Our study highlighted challenges in AI accountability, including inherent biases in LLM prompts that can skew results \cite{sharma2024generative} and the risk of confirmation bias in AI tool usage. We observed that participants without AI expertise struggled with effective prompting, often overgeneralizing or incorrectly equating AI interaction with human communication \cite{zamfirescu2023johnny}. This was reflected in participant behavior—P6 emphasized the importance of being ``proven wrong'' for learning, while P1 noted AI's tendency to be overly agreeable. Some participants (P9, P11) deliberately limited AI use to basic summarization to avoid potential bias in more complex analyses. This suggests tools need explicit features for bias identification and mitigation, helping PMs distinguish between objective data synthesis and potentially biased interpretations.

\subsubsection{AI Use Audit Trail}

Trust in today's workplace extends beyond human colleagues to encompass AI tools. In their high stakes
work environment, many PMs recognized the importance of delivering accurate and valuable work, both for the sake
of themselves and their peers. They valued knowing exactly where the data came from, which can be increasingly
difficult when handed materials and conclusions from other team members without additional context. P7's observation that LLM use often becomes a solitary activity also highlighted the risk of creating information silos \cite{woodruff2023knowledge}, particularly when AI tools become overly accommodating to individual viewpoints. Although research has examined transparency in AI algorithms for HR purposes \cite{park2022designing}, our findings emphasize the equal importance of transparency in AI interactions for PM tasks such as feature brainstorming. Thus, AI tools should incorporate lasting audit trails that allow for replicability. Audit trails must also go beyond tracking AI usage to actively support team alignment and stakeholder buy-in, both in a practical and motivational sense. Large organizations with multiple PMs benefit from knowledge sharing that provides context for AI outputs, allowing team members to reassess feature lists or discover new AI interaction techniques. This is particularly important given that LLMs can perform poorly with certain types of prompts, such as negated prompts \cite{jang2023can}. Beyond verification, audit trails build motivation and empathy among team members. Engineers better understand the importance of features through user feedback, while stakeholders gain insight into the rigorous brainstorming process that informs proposals.

\subsection{Enable Secure System Interoperability}

\textit{PM specific task:} Feature and bug fix prioritization

PMs must connect multiple systems and data sources to make informed decisions. Thus, when considering which features or bug fixes to prioritize first, PMs
need AI tools that balance the nuance of gathering and adapting to external information while safeguarding internal
information that is sensitive to the company. Our study reveals that effective PM work requires seamless yet secure integration between AI tools and existing company infrastructure, data pipelines, and business systems. 

\subsubsection{Security-Aware Integration}

AI tools must integrate securely with existing company systems while protecting sensitive information. Privacy emerged as a critical concern, with AI systems facing vulnerabilities including data poisoning and personal information leakage \cite{das2024security}. Our participants developed various coping strategies—some restricting themselves to company-internal chatbots, while others manually redacted sensitive information before using external tools like ChatGPT, which learns from user inputs \cite{gptprivacy}. Notably, PMs showed greater trust in AI tools integrated with familiar platforms like Slack, supporting research that known company logos increase perceived trustworthiness on unfamiliar sites \cite{lowry2005effect}. This suggests tools need built-in integration capabilities with approved company systems while maintaining clear boundaries for sensitive data handling. The current lengthy process of approving generative AI tools for corporate use creates significant adoption barriers. Tools need to provide clear security guarantees and integration paths to speed this process.

\subsubsection{Business Context and Priority Integration}

Tools must connect with existing business systems to support dynamic priority management and decision-making. As business
priorities shift, AI tools should allow PMs to iteratively refine and dig deeper on specific ideas or themes, enabling them
to explore different avenues. For instance, P3 asked the Q\&A widget for more details about pain points identified by the
Pain Point Tracker, and P11 broke down existing high level points in one Pain Point Tracker output by reusing them as
topics of focus for a new output of the Tracker. AI tools must support an interactive feedback loop for prioritization
\cite{tankelevitch2023metacognitive}, enabling PMs to delve deeper into feature or bug details to understand how they align with broader business goals
or evaluate trade-offs, such as when an engineer notes the high implementation cost of a large feature.
Yodeai's regenerable widgets demonstrated the value of allowing regenerability and maintaining consistent connections to underlying systems while supporting evolution in priorities over time.

\subsubsection{Multi-Pipeline Data Flow}

Tools must support intelligent integration of multiple data pipelines while maintaining data relevance and freshness. Our study revealed consistent desire for multiplatform data integration, particularly for B2C products where direct customer interviews may be impractical. Unlike general-purpose AI tools, PM-focused systems must enable explicit filtering and source differentiation across pipelines, as insight quality depends heavily on input data relevance. PMs expressed concerns about overwhelming AI systems with excessive or irrelevant data, which could
hinder the extraction of valuable insights. This concern is supported by existing research showing LLMs' sensitivity to irrelevant information in problem solving \cite{shi2023large} and RAG applications' susceptibility to distraction by semantically similar but irrelevant data \cite{wu2024easily}. The timeliness of data is also crucial, as PMs need up-to-date information
to assess product performance and evaluate the effectiveness of their actions. Several PMs noted their satisfaction with
the time-dependent graphs in the Pain Point Tracker. Therefore, AI tools that incorporate external data should include
auto-updating integrations or customizable weighting systems to prioritize certain data sets over others.

\subsection{Study Limitations}

Our study's depth-first approach, focusing primarily on PMs in software companies and specific tasks within Yodeai, provides valuable insights but may limit broader generalization. While PMs are representative of knowledge work domains and Yodeai's functionalities are applicable to various information synthesis tasks, replicating the study across diverse fields could validate and expand our findings. Participants noted that Yodeai's perceived usefulness depended on data scale and familiarity, suggesting that experiences might differ in production environments with more complex datasets. The short-term nature of our study, while offering in-depth initial insights, may not fully capture long-term benefits and challenges of generative AI in real-world settings. Despite these limitations, our research provides a crucial foundation for understanding immediate reactions to AI-assisted tools in knowledge work, informing both future research directions and the design of more comprehensive, longitudinal studies in this rapidly evolving field.

\subsection{Future Work}

Future research should expand upon our findings to further optimize AI-powered tools in knowledge work contexts. Longitudinal studies are crucial to investigate the long-term impact of tools like Yodeai on productivity, collaboration, and decision-making. Simultaneously, developing and evaluating strategies to mitigate potential risks and biases associated with AI use in knowledge work settings is essential, focusing on reducing confirmation bias, preventing overreliance on automation, and ensuring ethical deployment. Testing the effectiveness of our proposed design implications—adaptability, accountability, and interoperability—in real-world settings across diverse knowledge work domains will help refine and adapt these principles to better suit various professionals and industries. Additionally, exploring the potential applications of AI-powered tools beyond product management, such as in research, academia, journalism, and consulting, will provide valuable insights into the specific needs, challenges, and opportunities in these domains, leading to the development of more targeted and effective AI solutions for a broader range of knowledge work contexts.

\section{Conclusion}

Our research investigates the challenges and opportunities of integrating AI-powered tools in knowledge work, focusing on data structuring and decision-making processes. Through the development of Yodeai and user studies with knowledge workers and PMs, we uncovered the need for AI tools to allow for user control, verification methods, and collaborative approaches. Our findings highlight both the potential benefits of AI assistance in product management and critical limitations. We propose design implications prioritizing adaptability for diverse workflows, personal and collaborative accountability, and context-aware interoperability to address these challenges. As AI continues to evolve and integrate into professional settings, our work provides a foundation for developing user-centric AI tools that augment human capabilities rather than replace them. Moving forward, addressing the identified challenges and opportunities will be crucial in fostering effective human-AI collaboration, ultimately leading to more informed, efficient, and ethical decision-making processes across various knowledge work domains.

\bibliographystyle{ACM-Reference-Format}
\bibliography{6-refs, 6b-refs}

\end{document}